\documentclass[a4paper,11pt]{article}
\pdfoutput=1 % if your are submitting a pdflatex (i.e. if you have
\usepackage{jheppub} % for details on the use of the package, please
\usepackage[T1]{fontenc} % if needed

\usepackage{multirow} %

\title{\boldmath   A description of $\chi_{cJ}\to VV$ decays within the effective field theory framework}

 \author{Nikolay Kivel}
 \affiliation{ Physik-Department, Technische Universit\"at M\"unchen,\\ 
 James-Franck-Str. 1, 85748 Garching, Germany}

\abstract{
We study  $\chi_{cJ}\to VV$ decays using the QCD effective field theory approach. The helicity suppressed decay amplitudes are also  considered. The colour-singlet contributions of these amplitudes suffer from the endpoint singularities, it is shown that they  can be absorbed into renormalisation of  the nonfactorisable  colour-octet matrix element.  The latter can be associated with the colour-octet component of the charmonium wave function.  The heavy quark spin symmetry makes it possible  to establish the relationships  between colour-octet matrix elements for  different states $\chi_{cJ}$ up to higher order corrections in small velocity $v$. This allows us to estimate the polarisation parameters for $\chi_{c2}\to VV$ using data for $\chi_{c0,1}\to VV$.  This analysis is carried out for the available data on the  $\chi_{cJ}\to \phi\phi$ decays.}

\newcommand{\ba}{\begin{array}}
\newcommand{\ea}{\end{array}}

\newcommand{\Dsl}[1] { \setbox0=\hbox{$#1$}     
\dimen0=\wd0   \setbox1=\hbox{/} \dimen1=\wd1  \ifdim\dimen0>\dimen1        
 \rlap{\hbox to \dimen0{\hfil/\hfil}}  #1 \else \rlap{\hbox to \dimen1{\hfil$#1$\hfil}}  /  \fi  }
\newcommand{\bea}{\begin{eqnarray}}
\newcommand{\eea}{\end{eqnarray}}

\newcommand{\ns}{\Dsl{n}}
\newcommand {\nb}{\bar{n}}
\newcommand {\nbs}{\Dsl{\bar n}}
\newcommand{\nbn}{\frac{\nbs\ns}{4}}
\newcommand{\nnb}{\frac{\ns\nbs}{4}}

\newcommand{\bsd}{ \boldsymbol{\Delta}}

\begin{document} 
\maketitle
\flushbottom

\section{Introduction}
\label{sec:intro}
The branching fractions of $\chi_{cJ}$ decays into  different pairs of vector mesons like $\omega\omega$, $\phi\phi$ and $K^*\bar K^*$  have been measured in last two decades by BES \cite{BES:2005iaq,Ablikim:2006vm,BESII:2011hcd}  and BELLE \cite{Belle:2007qae,Belle:2012qqr} collaborations, see Table~\ref{tab1}.  The current  BES measurements are based on the accumulated 448 million $\psi(2S)$ decays \cite{BESIII:2020nme} that provided  access to  $\chi_{cJ}$ decays  through  radiative decays $\psi(2S)\to \chi_{cJ}+\gamma$.  The increasing of statistics  to several billion $\psi(2S)$ events by BESIII in future \cite{BESIII:2020nme}  opens up opportunities for a more accurate and detailed analysis of rare exclusive $\chi_{cJ}$ decays.  In particular, sufficiently accurate data can  provide a more detailed information  about the various helicity amplitudes, which describe  $\chi_{cJ}\to VV$ decays. Such information can be very useful for a better understanding of the underlying  hadronic dynamics in charmonium decays.  The first step in this direction has already been taken recently in the work \cite{BESIII:2023zcs}, where the helicity amplitudes of $\chi_{cJ}\to \phi\phi$ decays were studied for the first time.
\begin{table}[h!]
\centering
 \caption{Branching fractions for different decay channels $\chi_{cJ}\to VV$ in units $10^{-4}$.  The data are taken from  
 ref.~\cite{Workman:2022ynf}.  } 
 \label{tab1}
\begin{tabular}[c]{|c|c|c|c|}\hline
 Br & $\omega\omega$ & $\phi\phi$  & $K^{\ast}(892)^0\bar K^{\ast}(892)^0$\\\hline
$\chi_{c0}\to VV$ & $9.7\pm1.1$ & $8.0\pm0.7$ & $17\pm6$ \\\hline
$\chi_{c1}\to VV$ & $5.7\pm0.7$ & $4.2\pm0.5 $ & $14\pm4$ \\\hline
$\chi_{c2}\to VV$ & $8.4\pm1.0$ & $10.6\pm0.9$ & $23\pm 4$ \\\hline
\end{tabular}
\end{table}

Various approaches have been proposed to describe $\chi_{cJ}\to VV$ decay dynamics.  The  hadronic loop mechanism is considered in refs.~\cite{Chen:2009ah,Liu:2009vv,Huang:2021kfm},  perturbative QCD and  quark pair creation model in refs.~\cite{Zhou:2004mw,Chen:2013gka}.  The QCD factorisation framework based on the collinear factorisation  has been used  in ref. \cite{Chen:2012ih}.   However, these predictions can not describe  the available experimental data on  $\chi_{cJ}\to \phi\phi$  decays \cite{BESIII:2023zcs}.

In the present  work we continue to develop the QCD factorisation approach  using NRQCD \cite{Caswell:1985ui}  and soft-collinear effective theory frameworks (SCET) \cite{Bauer:2000yr,Bauer:2001yt,Beneke:2002ph,Beneke:2002ni}.  A special attention is devoted  to a description of the helicity suppressed amplitudes, which possess  the infrared (endpoint) divergencies \cite{Chen:2012ih}.  These  singularities  indicate about the violation  the collinear factorisation due to the mixing with the  colour-octet component of the charmonia wave function.  

 Existing data on $\chi_{cJ}$ hadronic decays indicate about large effects associated with the violation of  QCD helicity selection rule \cite{Brodsky:1981kj}.    As a rule, a description of  corresponding colour-singlet amplitudes involves the higher Fock states for the final state hadrons and therefore such contributions  are usually  suppressed by additional powers  of $1/m_c$ comparing to helicity conserving amplitudes.  In addition,  the contribution of the colour-octet component of the  charmonium wave function can mix with the colour-singlet contribution, which complicates a theoretical  description.  There is an evidence that  the octet contribution  provides a large numerical effect  and therefore  must be  included into the theoretical description.  
 
 Attempts to build a theoretical formalism for the colour-octet mechanism in  exclusive hadronic decays  were considered in refs.~\cite{Bolz:1996wh} for $\chi_{cJ}\to \pi\pi$ and in ref.\cite{Wong:1999dj} for $\chi_{cJ}\to p\bar p$ decays. However, suggested  approach represents a kind of a phenomenological model and can not be related with a systematic  effective field theory framework.  It is interesting to note that endpoint divergences are also one of the problematic features of this approach.  
 
 The mixing of the colour-octet and colour-singlet amplitudes in the context of  effective field theory  was  discussed in $B\to \chi_{cJ} K$-decays \cite{Beneke:2008pi} and $\chi_{cJ}\to \bar K K^*$ decays \cite{Kivel:2018rgd}.  It was demonstrated  that  the  endpoint divergences in the colour-singlet contribution  can be  absorbed into renormalisation of the colour-octet matrix element.  In ref.  \cite{Kivel:2018rgd} it is also shown that  the heavy quark spin symmetry allows to relate the octet matrix elements for different states $\chi_{cJ}$. This makes it possible to obtain a relationship between the  amplitudes with different spin $J$ and calculate the well defined spin symmetry breaking corrections.  The latter allows one  to get relations for different observables, which can be verified experimentally.  The goal of this work is to carry out a similar analysis for the amplitudes of  $\chi_{cJ}\to VV$ decays.

\section{ Decay amplitudes and observables }
In order to describe $\chi_{cJ}\to VV$ decays we define the following amplitudes 
\begin{equation}
\left\langle V(k, e_V)V(k^{\prime},e'_V)\right\vert i\hat T \left\vert \chi_{cJ}
(P,\epsilon_{\chi})\right\rangle =i(2\pi)^{4}\delta^{(4)}(P-k-k^{\prime})\ A_{J},%
\label{defAJ}
\end{equation}%
 The particle momenta satisfy on-shell relations
\begin{align}
k^2=k'^2=m_V^2,\   P^2=M_\chi^2.
\end{align}
where $m_V$ and $M_\chi$ denote the masses. For simplicity,  we  do not take into account the difference in masses of the  quarkonium states and assume $M_{\chi_{c0}}\simeq M_{\chi_{c1}}\simeq M_{\chi_{c2}}=M_{\chi}$.  For the  polarisation vectors of different  mesons we use short notations $e^\lambda_V(k)\equiv e_V$, $e^{\lambda'}_V(k')\equiv e'_V$, $\epsilon^{i}_{\chi}(P)\equiv \epsilon_{\chi}$.  The following normalisations relations are true 
\begin{align}
\sum_{\lambda} (e_V)_\mu (e_V)^*_\nu =-g_{\mu\nu}+\frac{k^\mu k^\nu}{m_V^2}, 
\quad 
\sum_{i} (\epsilon^i_{\chi})_\mu (\epsilon^i_{\chi})^*_\nu =-g_{\mu\nu}+\frac{P^\mu P^\nu}{M_\chi^2},
 \\ 
\sum_{i} (\epsilon^i_{\chi})_{\mu\nu} (\epsilon^i_{\chi})^*_{\alpha\beta}
=\frac{1}{2}M_{\mu\alpha}M_{\nu\beta}+\frac{1}{2}M_{\mu\beta}M_{\nu\alpha}-\frac{1}%
{3}M_{\alpha\beta}M_{\mu\nu},
\end{align}
where $M_{\alpha\beta}=\frac{P^{\alpha}P^{\beta}}{M_{\chi}^{2}}-g^{\alpha\beta}$.  

In order to decode the structure of the amplitudes $A_J$ defined in (\ref{defAJ}) it is convenient to introduce the following auxiliary tensors 
\begin{align}
 \ R_{T}^{\mu\nu}=g^{\mu
\nu}-\frac{1}{(k\cdot k^{\prime})^{2}-m_{V}^{4}}\left\{  (k\cdot k^{\prime})(k^{\mu
}k^{\prime\nu}+k^{\nu}k^{\prime\mu})-m_{V}^{2}\left(  k^{\mu}k^{\nu}%
+k^{\prime\mu}k^{\prime\nu}\right)  \right\},
\\
k_\mu R_{T}^{\mu\nu}=k'_\mu R_{T}^{\mu\nu}=0,\quad   V_{T}^{\mu}=R_{T}^{\mu\nu}V_{\nu}, \ R_{T}^{\mu\nu}g_{\mu\nu}=2,\ \ R_{T}^{\mu\nu}(R_{T})_\mu^{\phantom{\mu}\nu^{\prime}}=R_{T}^{\nu\nu^{\prime}}, 
\\
i\varepsilon\lbrack\mu,\nu]=\frac{2}{(k-k^{\prime})^{2}} i\varepsilon_{\mu\nu\alpha\beta}k^{\prime\alpha}k^\beta,\  i\varepsilon\lbrack k,\nu]=i\varepsilon\lbrack k',\nu]=0, \ i\varepsilon[V,\nu]\equiv i\varepsilon\lbrack\mu,\nu]V^\mu ,
\end{align}%
where $V_\mu$ denotes an arbitrary Lorentz vector and $i\varepsilon_{\mu\nu\alpha\beta}$ is Levi-Civita symbol. 

Using these definitions we define the following scalar amplitudes:
\begin{align}
A_{0}(i,\lambda,\lambda')&=\ \left(  e_{V}^{\ast}\cdot k^{\prime}\right)  \left(  e_{V}^{\prime \ast
}\cdot k\right)  \frac{4}{(k-k^{\prime})^{2}}\frac{m_{V}^{2}}{M_{\chi
}^{2}}A_{0}^{(LL)}+\left(  e_{V}^{\ast}\right)_T\cdot\left(e_{V}^{\prime\ast}\right)_T\ A_{0}^{(TT)},
\label{def:A0}
\\
A_{1}(i,\lambda,\lambda')&=\left\{  i\varepsilon\lbrack\epsilon_{\chi},e_{V}^{\ast}]\left(e_{V}^{\prime\ast}\cdot k\right)  -i\varepsilon\lbrack\epsilon_{\chi}%
,e_{V}^{\prime\ast}]\left(  e_{V}^{\ast}\cdot k^{\prime}\right)  \right\} \frac{m_{V}}{M^2_{\chi}}A_{1}^{(LT)},
\label{def:A1}
\\
A_{2} (i,\lambda,\lambda')&  =\epsilon_{\chi}^{\mu\nu}\ k_{\mu}k_{\nu}\left(  e_{V}^{\ast}\cdot
k^{\prime}\right)  \left(  e_{V}^{\prime\ast}\cdot k\right)  \frac
{16}{(k-k^{\prime})^{4}}\ \frac{m_{V}^{2}}{M_{\chi}^{2}}A_{2}^{(LL)}
\nonumber \\
&  +\epsilon_{\chi}^{\mu\nu}\left\{  k_{\mu}\left(  e_{V}^{\ast}\right)
_{\nu}^{T}\left(  e_{V}^{\prime\ast}\cdot k\right)  +k_{\mu}^{\prime
}\left(  e_{V}^{\prime\ast}\right)  _{\nu}^{T}\left(  e_{V}^{\ast}\cdot
k^{\prime}\right)  \right\}  \frac{4}{(k-k^{\prime})^{2}}\frac{m_{V}}{M_{\chi
}}A_{2}^{(LT)}
\nonumber  \\
&  +\epsilon_{\chi}^{\mu\nu}\left(  e_{V}^{\ast}\right)  _{(\nu}^{T}\left(
e_{V}^{\prime\ast}\right)  _{\mu)}^{T}\ A_{2t}^{(TT)}+\epsilon_{\chi}%
^{\mu\nu}\ k_{\mu}k_{\nu}\frac{4}{(k-k^{\prime})^{2}}
\left(  e_{V}^{\ast}\right)_T\cdot\left(  e_{V}^{\prime\ast}\right)_T%
\ A_{2s}^{(TT)},
\label{def:A2}
\end{align}
where  we use short notation for the symmetric traceless combination
\begin{equation}
\left(  e_{V}^{\ast}\right)  _{(\nu}^{T}\left(  e_{V}^{\prime\ast}\right)
_{\mu)}^{T}=\left(  e_{V}^{\ast}\right)  _{\nu}^{T}\left(  e_{V}%
^{\prime\ast}\right)  _{\mu}^{T}+\left(  e_{V}^{\ast}\right)  _{\mu}^{T
}\left(  e_{V}^{\prime\ast}\right)  _{\nu}^{T}-R_{T\mu\nu}\left(
(e_{V}^{\ast})^T\cdot (e_{V}^{\prime\ast})^T\right) .
\end{equation}
All scalar amplitudes $A_J^{(\dots)}$ have dimension of mass and can be associated with the different  helicity amplitudes  $\chi_{cJ}(i)\to V(\lambda) V(\lambda')$.   The coefficients in front of different Lorentz structures in Eqs.(\ref{def:A0})-(\ref{def:A2}) are chosen in order to get more simple analytical expressions for the corresponding decay rates $\Gamma[\chi_{cJ}\to VV]$.  The standard calculations give\footnote{We emphasize that these expressions are exact, i.e., they do not contain kinematic approximations. } 
 \begin{align}
& \Gamma[\chi_{c0}\to VV]=\left(  \frac{1}{2}\right)  ^{\delta_{VV}}%
\frac{1}{16\pi}\frac{\beta_{V}}{M_{\chi}}\left(  \ \left\vert A_{0}%
^{(LL)}\right\vert ^{2}+2\left\vert A_{0}^{(TT)}\right\vert ^{2}\right),
\label{Gam0}
 \\
& \Gamma[\chi_{c1}\to VV]=\left(  \frac{1}{2}\right)  ^{\delta_{VV}}%
\frac{1}{16\pi}\frac{\beta_V}{M_{\chi}}\frac{1}{3}\left\vert A_{1}%
^{(LT)}\right\vert ^{2},
\label{Gam1}
 \\
& \Gamma[\chi_{c2}\to VV]=\left(  \frac{1}{2}\right)  ^{\delta_{VV}}%
\frac{1}{16\pi}\frac{\beta_V}{M_{\chi}}\frac{1}{5}\left(  \frac{2}{3}\left\vert
A_{2}^{(LL)}\right\vert ^{2}+8\left\vert A_{2t}^{(TT)}\right\vert
^{2}+2\left\vert A_{2}^{(LT)}\right\vert ^{2}+\frac{4}{3}\left\vert
A_{2s}^{(TT)}\right\vert ^{2}\right),
\label{Gam2}
\end{align}
 where $\beta_V=\sqrt{1-4m_V^2/M^2_\chi}$ and $\delta_{VV}=1$ for identical states like $\omega\omega, \phi\phi$ and zero otherwise.

The superscripts $L$ and  $T$ correspond to longitudinally and transversely polarised mesons in the final state $VV$. Using the  various decays modes for the vector  resonances $V$ one can get information about vector meson decays with definite helicities $\chi_{cJ}\to V(\lambda)V(\lambda')$. The ratios of the different helicity  amplitudes can be extracted from the angular distributions of the cascade process $e^+e^-\to \psi(2S)\to \chi_{cJ}+\gamma\to (VV)+\gamma$ where the vector mesons decay into stable hadrons $V\to ff'$.  A more detailed  formalism for such helicity analysis was already considered in refs.\cite{Huang:2021kfm, Chen:2013gka, Chen:2020pia}.  The  helicity amplitudes $F^{J}_{\lambda,\lambda'}$ defined in refs.\cite{Huang:2021kfm, Chen:2013gka}  can be easily related to the scalar amplitudes defined in Eqs.(\ref{Gam0})-(\ref{Gam2}) and the corresponding formulas will be considered later. 

 The total widths  are given by  the sum of the different  partial rates
\begin{align} 
 &\Gamma[\chi_{c0}\to VV]=\Gamma_{0}^{(LL)}+\Gamma_{0}^{(TT)},\  \Gamma[\chi_{c1}\to \phi\phi]\equiv \Gamma_{1}^{(LT)},  
 \\
&\Gamma[\chi_{c2}\to VV]=\Gamma_{2}^{(LL)}+\Gamma_{2t}^{(TT)}+\Gamma_{2}^{(LT)}+\Gamma_{2s}^{(TT)},
\end{align}
where each term in the {\it rhs}  includes the appropriate amplitude from Eqs. (\ref{Gam0}) and (\ref{Gam2}).  
%In this work we consider these decay rates as independent observables and focus our attention on the theoretical description of the corresponding decay amplitudes.   

QCD analysis and  helicity conservation in the perturbative QCD interactions suggest the following hierarchy of helicity amplitudes
\begin{equation}
A_{0}^{(LL)}:A_{2}^{(LL)}:A_{2t}^{(TT)}:A_{1}^{(LT)}:A_{2}^{(LT)}:A_{0}^{(TT)}: A_{2s}^{(TT)}=1:1:1:\lambda:\lambda:\lambda^2:\lambda^2,
\label{hierarchy}
\end{equation}
where $\lambda\sim \Lambda/m_c$ and $\Lambda$ is the typical hadronic scale $\Lambda\sim \Lambda_{QCD}$.  The power suppressed amplitudes are sensitive to the higher Fock components of  meson wave functions. Using the terminology of the collinear factorisation, the nonperturbative coupling to mesons is described by the higher-twist operators that gives additional powers of the soft scale.  

  In the formal heavy quark limit $m_Q\to \infty$ the power suppressed amplitudes are expected to be small.  However, various  existing data indicate that the mass $m_c$ is not  large enough and effects from the helicity suppressed amplitudes are sufficiently large.   As an example, we  refer to the value of the  branching fraction  
  $\chi_{c1}$  in Table~\ref{tab1}.  Therefore a study of such amplitudes can be important  for a consistent  description of the data.

\section{The leading-power amplitudes $A_0^{LL}$, $A_2^{LL}$ and $A_{2t}^{TT}$}
In the limit $m_c\to \infty$  decay amplitudes are dominated by the colour-singlet contribution, which  can be  described  within the  collinear QCD factorisation framework, see {\it e.g.}  refs.~\cite{Lepage:1980fj, Chernyak:1983ej}.  Such  description allows one  to unambiguously separate the physics associated with the  short- and long-distance dynamics. The short distance subprocess $c\bar c\to 2 g^*\to (q\bar q)(q\bar q)$ is described in pQCD and can be computed systematically order by order in $\alpha_s$.  Corresponding calculations are well known in the literature and therefore we skip the technical details in this section. 

 The nonperturbative dynamics is described by the various long-distance matrix elements defined in the collinear QCD and NRQCD. 
The collinear matrix elements are parametrised  in terms of the  light-cone distribution amplitudes (LCDAs).  To the leading-power  accuracy the required LCDAs are defined by the leading twist quark-antiquark operators on the light-cone $z_1^2=z_2^2=0$
 \begin{equation}
 \langle V(k', e_V)| \bar \chi_{n} (z_1)\Gamma \chi_{n}(z_2)|0\rangle, 
 \label{Otw2}
 \end{equation}
 where $\Gamma$ denotes some Dirac structure and  $\chi_n(z)$  denotes the field describing the quark jet with the momentum $\sim k'$, which is collinear to  light-like vector $n$, see more details in  Appendix~\ref{setup}.
 
 Corresponding  LCDAs  can also be obtained from Bethe-Salpeter wave functions at (almost) zero transverse separations of the constituents (assuming the valence quark-antiquark component of the wave function) 
 \begin{equation}
 \phi(x,\mu) \sim \int_{k_\perp< \mu} \psi(x,k_\perp) d^2k_\perp. 
 \end{equation}
  In Appendix~\ref{def:LCDA} we  provide the  definitions of various LCDAs, which are used in this work. 
  
  For the NRQCD matrix elements of $\chi_{cJ}$ states  we use standard definitions, see {e.g.} \cite{Bodwin:1994jh},  which corresponds
 \begin{align}
\frac{1}{3}| \left\langle 0\right\vert \chi\dag\left(-\frac{i}{2}\overleftrightarrow{\boldsymbol{D}}\cdot\boldsymbol{\sigma }\right)\psi
\left\vert \chi_{c0}\right\rangle |^2=\frac{9}{2\pi} R'_{21}(0)^2\equiv f^2_\chi,
\label{def:Rp21}
 \end{align}
 where $R'_{21}(0)$ denotes  the radial wave function at the origin.

The  expressions for the amplitudes can be written as follows
\begin{align}
A_{i}^{(LL)}  & =\sqrt{ 2M_\chi} \frac{f_{\chi}f_{V}^{2}}{m_{c}^{4}}\left(  \pi\alpha
_{s}\right)  ^{2}\frac{2}{27} J_{i}^{(LL)}, \   A_{2t}^{(TT)}  =\sqrt{ 2M_\chi} \frac{f_{\chi}f_{V}^{^\bot 2}}{m_{c}^{4}}\left(  \pi\alpha
_{s}\right)  ^{2}\frac{2}{27} J_{2t}^{(TT)}, 
\label{Ai**}
\end{align}%
where  the different dimensionless  convolution integrals are given by ($\bar x= 1-x$)
\begin{align}
J_{0}^{(LL)}  & =\frac{1}{\sqrt{3}} \int_0^1 dx\  \frac{\phi_{2V}^{\Vert}(x)}{x\bar x}
\int_0^1 dy\ \frac{\phi_{2V}^{\Vert}(y)}{y \bar y}\frac{1}{xy+\bar x\bar y  }\left(
\frac{\left(  x-\bar y\right) ^{2}}{xy+\bar x\bar y }-2\right) ,
\label{J0LL}
\\
J^{(LL)}_{2}&= \int_0^1 dx\ 
\frac{\phi_{2V}^{\Vert}(x)}{x\bar x}
\int_0^1 dy\
\frac{\phi_{2V}^{\Vert}(y)}{y \bar y }\frac{1}{xy+\bar x\bar y  }\left(  1-\frac{\left(  x-\bar y\right)  ^{2}
}{xy+\bar x\bar y }\right),
\label{J2LL}
\\
J^{(TT)}_{2t} &=\int_0^1 dx\ 
\frac{\phi_{2V}^{\bot}(x)}{x\bar x}
\int_0^1 dy\ 
\frac{\phi_{2V}^{\bot}(y)}{y\bar y}\frac{1}{xy+\bar x\bar y  }.
\label{J2tTT}
\end{align}%
The soft couplings $f_V$ and $f^\bot_V$ have dimension of mass.   All the integrals over the momentum fractions $x$ and $y$ in Eqs.(\ref{J0LL})-(\ref{J2tTT}) are well defined because the LCDAs have smooth endpoint behaviour  $\phi_{2V}^{\Vert,\bot}(x)\sim x\bar x$ that compensates the singularities in the integrand denominators.  The running QCD coupling $\alpha_s$ is defined at the scale $\mu\sim m_c$.  A more detailed discussion of the various parameters and the models for the LCDAs are considered  in  Appendix~\ref{def:LCDA}.

\section{Helicity suppressed amplitudes}

Collinear  factorisation for helicity suppressed  amplitudes is violated by the  endpoint divergencies in the collinear integrals ref.~\cite{Chen:2012ih}. Such divergencies indicate about the overlap  of  collinear and  ultrasoft (usoft) modes.  The latter describes the particles with $p_{us}\sim m_cv^2\sim \Lambda$.  Below it will be shown  that the structure of the endpoint divergencies  is closely associated with the  configuration where one of the gluons in the annihilation subprocess $c \bar c\to g^* g^*\to (\bar q q)  (\bar q q)$ becomes usoft: $g^* g^*\to g_{us}^* g_h^*$.  Therefore the subprocess with the two hard gluons ($g^*_h g^*_h$) transforms  into  the subprocess  with the one hard gluon $c \bar c\to g_h^*\to (\bar q q)$.  Such  subprocess is closely associated with the annihilation of the colour-octet configuration of $c\bar c$ pair.       The endpoint divergencies indicate about the overlap  of  colour-singlet and colour-octet contributions.

In such a situation, the  factorisation is  described by the sum of  colour-singlet  and the  colour-octet  LDMEs. Let us consider, as example, the amplitude $A_1$.  Schematically,  the  factorisation formula for this amplitude  can be written as 
\begin{equation}
A^{(LT)}_{1}= \phi_{2V}\ast T_{1}\ast \phi_{3V} + C_{h}  a^{(oct)}_{1} ,
\label{A1LT}
\end{equation}
where the first term in the {\it rhs} denotes colour-singlet collinear integral with the hard kernel $T_1$ and  twist-2 and  twist-3 LCDAs denoted as $ \phi_{2,3V}$. The hard kernel $T_1$ corresponds to the hard transition   $c \bar c\to g_h^* g_h^*\to (\bar q q)  (\bar q q)$.    The second term in the {\it rhs} of Eq.(\ref{A1LT})  includes the  coefficient function $C_h$ associated with  the hard subprocess $c \bar c\to g_h^*\to (\bar q q)$  and the  octet  nonfactorisable LDME
\begin{equation}
  a^{(oct)}_{1} \sim  \left\langle VV \right\vert    \mathcal{O}_{oct}  \left\vert \chi_{c1}\right\rangle.
\label{aoct}
\end{equation}
The  operator $\mathcal{O}_{oct}\sim (c \bar c)_8 (q\bar q)_8$ consists of  the two colour-octet components, which are constructed from the NRQCD fields $(c \bar c)_8$ and from the hard-collinear fields in SCET-I $(q\bar q)_8$.  The LDME  $a^{(oct)}_{1}$  is nonfactorisable because of usoft  exchanges between the initial and final states. 
The formula (\ref{A1LT}) implies that the  infrared endpoint singularities in the collinear  convolution $\phi_{2V}\ast T_{1}\ast \phi_{3V}$ can be  absorbed into the renormalisation of the  octet contribution $C_{h}  a^{(oct)}_{1}$.   

The described picture makes it possible  to develop  the  following interpretation.  The charmonium colour-octet configuration can be viewed as a compact $c\bar c$ colour-octet pair of size $\sim 1/(m_c v)$ surrounded by an ultrasoft  cloud of size $\sim 1/\Lambda$.  Such a configuration is in some sense similar to heavy-light systems, such as  $B$-mesons. Following to this analogy,  the operator $ \mathcal{O}_{oct}$  in the LDME  in Eq.(\ref{aoct}) plays the role of an external source, similar to the low-energy effective Lagrangian in the Standard Model.  This source instantly annihilates the colour-octet $c\bar c$ pair into light energetic quark-antiquark pair $(q \bar q)_8$.  This quark-antiquark pair hadronises into a final hadronic state interacting with the usoft  cloud from the parent charmonium.   

The similar factorisation formula as in Eq.(\ref{A1LT}) can also be written for the decay  $\chi_{c2}$  into the same final state $VV$, i.e.  for $A_2^{LT}$ amplitude
\begin{equation}
A^{(LT)}_{2}= \phi_{2V}\ast T_{2}\ast \phi_{3V} + C_{h}  a^{(oct)}_{2} ,
\label{A2LT}
\end{equation}
where the octet LDME is defined analogously $a^{(oct)}_{2} \sim  \left\langle VV \right\vert    \mathcal{O}_{oct}  \left\vert \chi_{c2}\right\rangle$. Keeping in mind the analogy with the heavy-light system noted above, one can expect that defined   nonfactorisable LDMEs  do not depend on the quantum numbers of the heavy $c\bar c$-pair in the limit  $m_c\to \infty$.  In particular, the ultrasoft modes  can not resolve the total angular momentum $J_{c\bar c}$ of the heavy quark-antiquark pair and  therefore the nonperturbative  dynamics does not depend on $J_{c\bar c}$.  These arguments  leads to an idea of approximate heavy quark spin symmetry (HQSS). This symmetry  make it possible  to get  the following relation  between the nonfactorisable  LDMEs  
\begin{equation}
 a^{(oct)}_{2}  =\lambda_{12}\ a^{(oct)}_{1} +\mathcal{O}(v^2),
\label{def:lam12}
\end{equation}
where $\lambda_{12}$ is some  number.    
  
  The Eq.(\ref{def:lam12}) can be used in order to establish relationship between the physical amplitudes.  Excluding from the expressions   (\ref{A1LT}) and  (\ref{A2LT}) the nonfacrorisable  LDMEs, with the help of Eq.(\ref{def:lam12}), one finds 
\begin{equation}
A_{2}^{(LT)}=\lambda_{12} A_{1}^{(LT)}
+ \phi_{2V}\ast \left(T_{2}-\lambda_{12} T_{1}\right)\ast \phi_{3V},
\label{A2A1T}
\end{equation}
where the second term on {\it rhs} is  well defined and represents the HQSS breaking  correction, which occurs due to the interactions at short distances due the subprocess  $c \bar c\to g^* g^*\to (\bar q q)  (\bar q q)$.  The infrared endpoint singularities must cancel in the combination $T_{2}-\lambda_{12} T_{1}$, hence  this correction can be calculated in collinear factorisation scheme.  The relation (\ref{A2A1T}) can be used  in order to restrict  the ratio $\Gamma_2^{(LT)}/\Gamma_1^{(LT)}$, which can be measured experimentally.  

The program described above is close in spirit to the well-known approach in  decays of B mesons \cite{Beneke:2000wa}. This scheme  was used  for description of  $\chi_{cJ}\to K \bar K^*$ decays in ref.~\cite{Kivel:2018rgd}.   In the following   we use  the similar  approach  for  helicity suppressed amplitudes in $\chi_{cJ}\to VV$ decays.

\subsection{Hard kernels and HQSS relation for  amplitudes  $A_1^{LT}$ and $A_2^{LT}$}

Let us start our analysis from the calculations of  hard kernels $T^{(LT)}_{1,2}$ describing the colour-singlet contribution.  At the next step,  we will analyse  the structure of the emerging IR-singularities and propose the additional contribution that need to be added to  the  factorisation formula.  

  To calculate the amplitudes $A_{1,2}^{(LT)}$,  various  twist-3 LCDAs are needed.\footnote{ We assume  the kinematical twist, which must be distinguished from the {\it geometrical} twist defined as dimension minus spin.  }  The  set of the corresponding twist-3 light-cone operators can be divided  on  valence (quark-antiquark) and non-valence (quark-antiquark-gluon) operators. In this work we consider  only the contributions associated with the valence operators and neglect the non-valence ones, for simplicity.  Schematically, the matrix elements of the valence operators can be described as 
\begin{equation}
 \langle V(k') \vert  \bar \chi_n (z_1) \Gamma \partial_\bot \chi_n(z_2)  \vert 0\rangle,  \, z_i^2=0, 
 \label{Otw3}
\end{equation}
where  $\partial_\bot$ is the transverse derivative. The presence of the derivative provides additional power of the small scale  $\lambda\sim \Lambda/m_c$,  and the corresponding operators have the kinematical twist-3.  This transverse derivative can be associated with the quark-antiquark angular momentum $L=1$ as required by the total angular momentum conservation in the hard subprocess.    The full set of required valence matrix elements and their parametrisation in terms of LCDAs are  presented in Appendix~\ref{def:LCDA}.  

 The  Lorentz symmetry and QCD equations of motions make it possible  to derive  the valence twist-3 LCDAs  in terms of the  twist-2 LCDAs $\phi_{2V}$. Such an approximation is also also known in literature as Wandzura-Wilczeck one. It has been verified  in many various calculations that  twist-3 valence  contributions provide a consistent theoretical description preserving all basic symmetries of the QCD.  
 
 The analytical expression for the colour-singlet contributions to the amplitudes $A_{1,2}$  can be written as 
 \begin{align}
 \left.A_J\right |_{cs} = & \frac{2}{27}(ig)^4 \sqrt{2M_\chi}f_\chi \int_0^1dx \int_0^1 dy \  
 \text{Tr}[\gamma^\alpha \hat M_3(x, k) \gamma^\beta \hat M_2(y,k')+\gamma^\alpha \hat M_2(x,k) \gamma^\beta \hat M_3(y,k')]
 \nonumber \\
 &\times \frac{(-i)}{(k_1+k'_2)^2}\frac{(-i)}{(k_2+k'_1)^2}
  \frac{1}{4}\text{Tr}\left[ \mathcal{P}_{J}^{\mu} \left(  \frac{1}{2m_c}\left\{  \gamma^{\mu},D^{\alpha\beta}(k_i,k'_i)\right\}  +D^{\prime\alpha\beta\mu }(k_i,k'_i)\right)  \right].
  \label{calc:AJLT}
 \end{align}
 For a convenience,  we define  the prefactor consisting of the colour factor $C_F/(2N_c^2)=2/27$ and the vertex factors $(ig)^4$, the multiplier $\sqrt{2M_\chi}f_\chi$  is associated with the  NRQCD matrix element.   
 
\begin{figure}[h!]
\centering % \begin{center}/\end{center} takes some additional vertical space
\includegraphics[width=4in]{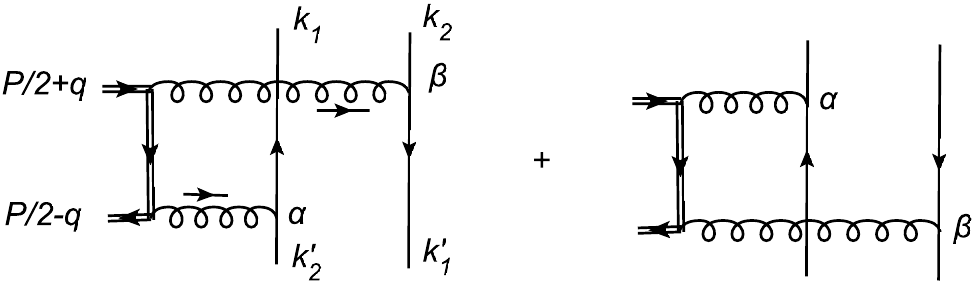}
\caption{\label{pQCD_graphs} The pQCD diagrams, which describe the hard kernels in the decay amplitudes. The heavy quark momenta (double lines) are incoming and all the light quark  momenta are outgoing.  The relative quark momentum $q$ in Eqs.(\ref{Dab}) and  (\ref{Dabprime}) is set to zero after differentiation. }
\end{figure}

 The Dirac trace in the first line (\ref{calc:AJLT}) describes the light quark subdiagram with the twist-2  and  twist-3 projections $\hat M_{2,3}$
  for the light-cone matrix elements and the matrices $\gamma^{\alpha, \beta}$ from the QCD vertices, see Fig~\ref{pQCD_graphs}.
  
   The second line in (\ref{calc:AJLT}) includes the two gluon propagators (in Feynman gauge) and the trace involving the Dirac matrices from the heavy quark line. The outgoing collinear  quarks and antiquarks  are defined with  the following momenta
 \begin{align}
k_{1}=x_{1} k+k_{\bot},\ k_{2}=x_{2} k- k_{\bot},\ k=(kn)\bar n/2,  \ x_1\equiv x, x_2 \equiv \bar x,
\label{k12}
\\ k_{1}^{\prime}=y_{1}k^{\prime}+k'_{\bot},\, k_{2}^{\prime}=y_{2}k^{\prime}-k'_{\bot}, 
 k'=(k'\bar{n}) n/2, \ y_1\equiv y, y_2 \equiv \bar y.
 \label{y12}
\end{align}
The transverse components must be treated carefully because the twist-3 projections $M_3$ in the first line (\ref{calc:AJLT}) involve derivatives with the respect to  $k_{\bot}$ and  $k'_{\bot}$, which act on the analytical expression in the second line.  The explicit expressions for the different projections $M_i$ are given in Appendix~\ref{def:Projections}.

The heavy quark trace is described by the heavy quark subdiagram, it involves the  differentiation with respect to relative heavy quark momentum that gives  two terms  ($P=k+k'$)
\begin{equation}
(-i)D^{\alpha\beta}=\frac{\gamma^{\alpha}\left(  - \Dsl{P}/2+ \Dsl{k}_{1}+ \Dsl{k}_{2}^{\prime}+m_c\right)  \gamma^{\beta}}{  \left(  -P/2+k_{1}+k_{2}^{\prime}\right)^{2}-m_c^{2}+i0  }
+\frac{\gamma^{\beta}\left(   \Dsl{P}/2- \Dsl{k}_{1}- \Dsl{k}_{2}^{\prime}+m_c\right)
\gamma^{\alpha}}{  \left(  P/2-k_{1}-k_{2}^{\prime}\right)  ^{2}-m_c^{2}+i0  }.
\label{Dab}
\end{equation}%
\begin{align}
(-i)D^{\prime\alpha\beta\mu}&=\frac{\gamma^{\alpha}\gamma^{\mu}\gamma^{\beta}%
}{  \left(  {P}/2-k_{1}-{k}_{2}^{\prime}\right) ^{2}-m_c^{2}
}-2\left(   \Dsl{k}_{1}+ \Dsl{k}_{2}^{\prime}\right)  ^{\mu}\frac{\gamma^{\alpha}\left(
- \Dsl{P}/2+ \Dsl{k}_{1}+ \Dsl{k}_{2}^{\prime}+m_c\right)  \gamma^{\beta}}{\left[  
\left(  {P}/2- {k}_{1}- {k}_{2}^{\prime}\right)^{2}-m_c^{2}\right]  ^{2}}
\nonumber \\
&+\frac{\gamma^{\beta}\gamma^{\mu}\gamma^{\alpha}%
}{  \left(  {P}/2- {k}_{1}-{k}_{2}^{\prime}\right) ^{2}-m_c^{2}
}+2\left(   \Dsl{k}_{1}+ \Dsl{k}_{2}^{\prime}\right)  ^{\mu}\frac{\gamma^{\beta}
\left( \Dsl{P}/2- \Dsl{k}_{1}- \Dsl{k}_{2}^{\prime}+m_c\right)  \gamma^{\alpha}}{\left[  
\left( {P}/2-k_{1}-{k}_{2}^{\prime}\right) ^{2}-m_c^{2}\right]  ^{2}}.
\label{Dabprime}
\end{align}
The Dirac projectors $\mathcal{P}_{J}^{\mu}$  for different charmonium states with $J=1,2$ are given by
\begin{equation}
\mathcal{P}_{1}^{\mu}=-\frac{1}{2\sqrt{2}}(1+\Dsl{ \omega})\left[  \gamma_{\top}^{\mu
},\Dsl \epsilon_\chi\right]  \gamma_{5},\  \mathcal{P}_{2}^{\mu}=\epsilon^{\mu\nu}_\chi(1+\Dsl \omega)\gamma_{\top}^{\nu},
\label{def:PJ}
\end{equation}
where $\omega$ denotes the velocity  $P=M_\chi \omega$, $ \omega^2=1$ and 
\begin{equation}
\gamma_{\top}^{\nu}=\gamma^{\nu} - \omega^\nu \Dsl{ \omega}.
\end  {equation}

Performing the calculation of the traces and contractions of the indices in Eq.(\ref{calc:AJLT}) and matching to the definitions in Eqs.(\ref{def:A1}) and (\ref{def:A2})  gives the following results
\begin{equation}
\left. A_{J}^{(LT)}\right|_{cs}=\sqrt{2M_\chi} \frac{f_{\chi} m_{V}}{m^3_{c}} \left(
\pi\alpha_{s}\right)  ^{2}\ \frac{2}{27}
\left\{ 
 \frac{f_{V}^{2}}{m_{c}^{2}} J^{(J)}_{V\Vert}+\frac{f_{V}^{\bot 2}}{m_{c}^{2}} J^{(J)}_{V\bot}
\right\},
\label{ALTcoll}
\end{equation}
where the  convolution integrals  can be conveniently written as
\begin{align}
 J^{(J=1)}_{V\Vert}&=-\frac{1}{\sqrt{2}}\left( J^{(s)}_{V\Vert} + J^{(r)}_{V\Vert} \right), \ 
 J^{(J=1)}_{V\bot}=-\sqrt{2}  J^{(s)}_{V\bot},
 \label{J1s}
 \\
 J^{(J=2)}_{V\Vert}&=\frac 12 J^{(s)}_{V\Vert},   \  J^{(J=2)}_{V\bot}= J^{(s)}_{V\bot},
 \label{J2s}
\end{align}
 with
 \begin{equation}
J^{(s)}_{V\bot}=\int_0^1 dx\ \frac{G_{3V}^{\Vert}(x )}{x\bar x}
\int_0^1 dy\ \frac{\phi^{\bot}_{2V}(y)}{y \bar y}\frac{1}{x y +\bar x \bar y}\left(  \frac{1}%
{x y +\bar x \bar y}-1+\frac{x-\bar y}{x y +\bar x \bar y}\right)  .
\label{JsVbot}
\end{equation}%
\begin{align}
J^{(s)}_{V\Vert} &  =\int_0^1 dx\ \frac{1}{x\bar x}\left\{  \left(  \bar x-x\right)  G_{3V}^{\bot}(x)-\tilde G_{3V}^{\bot}(x) \right\}  
\int_0^1 dy\ \frac {\phi^{\Vert}_{2V}(y)}{y \bar y}\frac{1}{(x y +\bar x \bar y)^{2}},
\label{JsVVert}
\\
J^{(r)}_{V\Vert} &  =\int_0^1 dx\ \frac{\tilde G_{3V}^{\bot}(x)}{x\bar x}
\int_0^1 dy\ \frac{\phi^{\Vert}_{2V}(y)}{y \bar y}\frac{8}{x y +\bar x \bar y}.
\label{JrVVert}
\end{align}%
Remind, the twist-3 LCDAs $G_{3V}^{\bot}, G_{3V}^{\Vert}, \tilde G_{3V}^{\Vert}$ describe the different valence light-cone matrix elements as in Eq. (\ref{Otw3}) and explicitly defined in Appendix~\ref{def:LCDA}.

 The integrals with the superscripts $(s)$ and $(r)$ are singular or regular, respectively.  The occurring  divergencies  are  closely associated with the  singularities of the hard kernels in the integrands.  From the structure of the integrands in Eqs.(\ref{JsVbot})-(\ref{JrVVert}) it follows that the singular behaviour is associated with the endpoint limits  $x\to 0(1)$ and $y\to 1(0)$, respectively.  In the collinear region  the momentum fractions are considered to be of order one $x\sim \mathcal{O}(1)$, which implies, see for instance  Eq.(\ref{k12}),   that $x (kn)\sim x m_c\sim m_c$.  However, if the collinear region overlaps with the usoft one, then the integration domain  with the small fractions $x\sim \lambda\sim \Lambda/m_c$ is not  suppressed by  $\lambda$ and  this usually  leads to  logarithmical divergencies in the collinear convolutions.  Therefore  our  task is to inspect the regions where  $x (\bar x) \sim \lambda$ and $y(\bar y ) \sim \lambda$ and to find the configurations of particle momenta  leading to  divergent integrals. 

In order to perform such analysis one has also to take into account  the properties of the various LCDAs.  The  different  twist-2 LCDAs, which enters into the integrals,  have the following endpoint behaviour
\begin{align}
\phi^{\Vert,\bot}_{2V}(y\to 0)=y[\phi^{\Vert,\bot}_{2V}]^\prime(0)+\mathcal{O}(y^2), 
\\
\phi^{\Vert,\bot}_{2V}(y\to 1)=-\bar y[\phi^{\Vert,\bot}_{2V}]^\prime(1)+\mathcal{O}(\bar y^2),
\end{align}        
where the prime denotes the derivative with respect to the argument $f'(x)=d/dx f(x)$. This behaviour is closely associated with the properties of the evolution kernels for these LCDAs.  

The twist-3 LCDAs in the valence approximation have the following  endpoint behaviour
\begin{align}
G_{3V}^{\bot}(x\rightarrow0) &=x [G_{3V}^{\Vert}]'(0)+\frac{x^{2}}{2}\frac{3}{2}  [\phi^{\Vert}_{2V}]^\prime (0)+\mathcal{O}(x^3),
\label{G3VIIu0}
\\
G_{3V}^{\bot}(x\rightarrow1)&=-\bar x [G_{3V}^{\Vert}]'(1)+\frac{\bar{x}^{2}}{2}\frac{3}{2}[\phi^{\Vert}_{2V}]^\prime(1)+\mathcal{O}(\bar x^3).
\label{G3VIIu1}
\end{align}
\begin{align}
\tilde G_{3V}^{\bot}(x\rightarrow0)&=x [\tilde G_{3V}^{\Vert}]'(0)- \frac{x^{2}}{2}[\phi^{\Vert}_{2V}]^\prime (0)+\mathcal{O}(x^3),
\label{tG3VIIu0}
\\
\tilde G_{3V}^{\bot}(x\rightarrow1)&=-\bar x [\tilde G_{3V}^{\Vert}]'(1) +\frac{\bar{x}^{2}}{2} [\phi^{\Vert}_{2V}]^\prime(1)+\mathcal{O}(\bar x^3).
\label{tG3VIIu1}
\end{align}
\begin{align}
G_{3V}^{\Vert}(x\rightarrow0)=x[G_{3V}^{\Vert}]'(0)+\mathcal{O}(x^2),\ \
G_{3V}^{\Vert}(x\rightarrow1)=-\bar{x} [G_{3V}^{\Vert}]'(1) +\mathcal{O}(\bar x^2).
\label{G3Vbot-x0}
\end{align}
The explicit expressions for the first derivatives in these formulas read
\begin{align}
[G_{3V}^{\bot}]'(0)&=-[\tilde G_{3V}^{\bot}]'(0)=-\frac{1}{2}\int_{0}^{1}dv\frac{\phi^{\Vert}_{2V}(v)}{v},  
\label{G3VIIx0}
\\
[G_{3V}^{\bot}]'(1)&=[\tilde G_{3V}^{\bot}]'(1)=-\frac{1}{2}\int_{0}^{1}dv\frac{\phi^{\Vert}_{2V}(v)}{1-v},  
\\
[G_{3V}^{\Vert}]'(0)&=-4\int_{0}^{1}dv\frac{\phi^{\bot}_{2V}(v)}{v},\ \ [G_{3V}^{\Vert}]'(1)=4\int_{0}^{1}dv\frac{\phi^{\bot}_{2V}(v)}{1-v}.
\end{align}
 The higher order terms shown in Eqs.(\ref{G3VIIu0})-(\ref{tG3VIIu1}) will be required later. 

Let us regularise the integrals in Eq.(\ref{JsVbot})-(\ref{JrVVert}) setting infrared cut-off  $\delta_{IR}<x,y$ and $\delta_{IR}<\bar x,\bar y$\footnote{The  fractions $x,\ y$ correspond to the quark fractions, see Fig.~\ref{pQCD_graphs}, while $\bar x,\bar y$ to antiquarks. },  which helps to  work  with the divergent integrals.  Consider as example  the region $\delta_{IR}<x <\eta_{UV}$ and $\delta_{IR} <\bar y< \eta_{UV}$, where the  parameter $\eta_{UV}$ plays the role of the  boundary, which separates the usoft region $x,\bar y\sim \lambda$ from the collinear region $x,\bar y\sim 1$.  Obviously, it is assumed that  $\delta_{IR}\ll \eta_{UV}$.  Expanding  the integrand with respect to small fractions $x$ and $\bar y$ and keeping the leading term  one finds
\begin{align}
[J^{(s)}_{V\bot}]_{us(x\sim \bar y\sim\lambda)} &= \int_{\delta_{IR}}^{\eta_{UV}} dx\ \frac{G_{3V}^{\Vert}(x)}{x\bar x}
\int_{\delta_{IR}} ^{\eta_{UV}} d\bar y\ \frac{\phi^{\bot}_{2V}(y)}{y \bar y}\frac{1}{x y +\bar x \bar y}\left(  \frac{1}%
{x y +\bar x \bar y}-1+\frac{x-\bar y}{x y +\bar x \bar y}\right)
\\
&=  \int_{\delta_{IR}}^{\eta_{UV}} dx\ [G_{3V}^{\Vert}]'(0)
\int_{\delta_{IR}}^{\eta_{UV}} d\bar y\ \frac{-[\phi^{\bot}_{2V}(1)]'}{(x  + \bar y)^2} +\mathcal {O}(\lambda).
\label{JsVbotUS}
\end{align}%
 The leading order usoft integral scales as $[J^{(s)}_{V\bot}]_{us}\sim \mathcal{O}(\lambda^0)$ because $dx\sim d\bar y\sim \lambda$, $(x  + \bar y)^{-2}\sim \lambda^{-2}$. Simple calculation yields the logarithm
 \begin{align}
 [J^{(s)}_{V\bot}]_{us(x\sim \bar y\sim\lambda)}&=- [G_{3V}^{\Vert}]'(0)[\phi^{\bot}_{2V}(1)]' \ln[\eta_{UV}/\delta_{IR}]+\mathcal {O}(1).
 \end{align}
 The similar  consideration  gives 
 \begin{align}
[J^{(s)}_{V\Vert}]_{us(x\sim \bar y\sim\lambda)} &
 \simeq \int_{\delta_{IR}}^{\eta_{UV}} dx  \left\{   [G_{3V}^{\bot}]'(0)-[\tilde G_{3V}^{\bot}]'(0) \right\}  
  \int_{\delta_{IR}}^{\eta_{UV}}  d\bar y\ \frac {-[\phi^{\Vert}_{2V}]'(1)}{(x  + \bar y)^{2}}
  \label{JsVVertUS1}
  \\[2mm]
&=-\left\{   [G_{3V}^{\Vert}]'(0)-[\tilde G_{3V}^{\Vert}]'(0) \right\}[\phi^{\Vert}_{2V}]'(1)\ln[\eta_{UV}/\delta_{IR}]+\mathcal {O}(1),
\label{JsVVertUS}
\\
[J^{(r)}_{V\Vert}]_{us(x\sim \bar y\sim\lambda)} &  \sim \mathcal{O}(\lambda).
\end{align}%

The similar results can be obtained  for the symmetric usoft region $\bar x\sim y\sim \lambda$
\begin{align}
[J^{(s)}_{V\bot}]_{us(\bar x\sim  y\sim\lambda)}&=- [G_{3V}^{\Vert}]'(1)[\phi^{\bot}_{2V}(0)]' \ln[\eta_{UV}/\delta_{IR}]+\mathcal {O}(1)
\\
[J^{(s)}_{V\Vert}]_{us(\bar x\sim  y\sim\lambda)}&=\left\{   [G_{3V}^{\bot}]'(1)+[\tilde G_{3V}^{\bot}]'(1) \right\}[\phi^{\Vert}_{2V}]'(0)\ln[\eta_{UV}/\delta_{IR}]+\mathcal {O}(1),
\\
[J^{(r)}_{V\Vert}]_{us(\bar  x\sim y\sim\lambda)} &  \sim \mathcal{O}(\lambda).
\end{align}
 All other possible endpoint regions  give  power suppressed contributions and therefore  can not be associated with the logarithmic endpoint singularities.  This also can be seen subtracting  the  usoft  contributions  from the regularised integrals, the results will not depend on the parameter $\delta_{IR}$.   
 
 The  contributions, which appear from the endpoint regions,  can be associated with the different  factorisation scheme, which involves usoft modes.  Such contributions must be included into the description along  with  the collinear singlet contribution described  in Eq.(\ref{ALTcoll}).  The structure of these usoft contributions can be understood  from the analysis of  the endpoint regions in the diagrams. Consider again the usoft limit  $x\sim \bar y\sim\lambda$,  then  the  external  quark with momentum $k_1$ and antiquark with momentum $k'_2$ can be understood as the  usoft  particles. Similarly the virtual gluon  producing this pair, see Fig.~\ref{pQCD_graphs}, is also usoft
 \begin{equation}
 (k_1+k'_2)^2 \simeq 4 x\bar y  m^2_2\sim \lambda^2 m_c^2\sim \Lambda^2.
 \end{equation}
 At the same time the second gluon remains hard
 \begin{equation}
 (k_2+k'_1)^2 \simeq 4\bar x y  m^2_2\simeq  4m_c^2.
 \end{equation}
 
 Emission of the usoft gluon does not change the virtuality of the potential heavy quark $p^2_Q\sim (m_c v)^2$,  therefore in the usoft  regions  the  momentum of the  heavy quark, see Fig.~\ref{pQCD_graphs}, remains unchanged 
 \bea
(P/2-k_{us})^2-m_c^2\sim (P\cdot k_{us})\sim m_c\Lambda \sim (m_c v)^2,
 \eea  
 which implies that $\Lambda \sim m_c v^2$ as it was assumed. 

\begin{figure}[h!]
\centering % \begin{center}/\end{center} takes some additional vertical space
\includegraphics[width=4in]{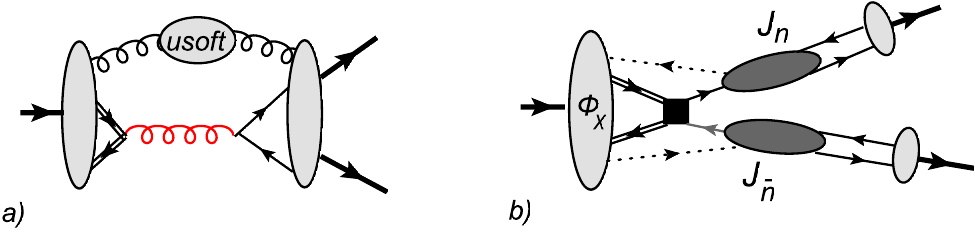}
\caption{\label{npQCD_graph} $a)$ The non-factorisable contribution, associated with the colour-octet amplitude. The red gluon line denotes the hard gluon with  momentum $p_h^2\sim m_c^2$. $b)$ A proposal for factorisation of the colour-octet matrix element in the limit $m_Q\to \infty$. The black square represent the colour-octet operator $\mathcal{ O}_8$, the  dotted fermion lines denote the usoft light quarks. The dark blobs denote the hard-collinear jet functions associated with the momenta $p^2_{hc}\sim \Lambda m_Q$, the  light-grey blobs denote the different  LCDAs.  }
\end{figure}

The similar  picture also holds for the  usoft limit $\bar x\sim  y\sim\lambda$. Therefore the hard subprocess in this case is associated with the  annihilation of the octet $Q\bar Q$ pair into light collinear-anticollinear  quark-antiquark  pair: $c\bar c\to g^* \to q\bar q$. The created  light quark and antiquark  interact with the usoft particles in order to hadronise into the hadronic final state.  Due to the  the soft-overlap between the initial and final states, see Fig.~\ref{npQCD_graph}$\ a)$,  corresponding matrix element  are not factorised after integration over the hard modes. 
 Factorising the hard modes  gives the four-fermion operator, which  can be represented as product of two colour-octet operators,  which are built from NRQCD  and SCET-I {\it  hard-collinear}  fields $\chi_{n,\bar n}$
 \bea
C_h\mathcal{ O}_{oct}=C_h \chi^{\dagger }T^a\sigma^\mu_{\top}\psi \ (\bar \chi_nT^a\gamma^\mu_\bot\chi_{\bar n}+\bar \chi_{\bar n}T^a\gamma^\mu_\bot\chi_{n}).
\label{def:Ooct}
 \eea
 where we assume that $\sigma^\mu=\{1,\boldsymbol{\sigma}\}$.  The  hard coefficient function $C_h = \pi\alpha_s/m^2_c$ is independent of the hadronic states. The  resulting  matrix elements 
 \bea
  \left\langle VV \right\vert  C_h   \mathcal{O}_{oct}  \left\vert \chi_{cJ}\right\rangle 
  \label{meOoct}
 \eea
still depends on the mass $m_c$ but this dependence is defined by the interactions of  hard-collinear particles with momenta  $p^2_{hc}\sim m_c \Lambda$.  Since the charm mass is not sufficiently large, the realistic hard-collinear scale $ m_c \Lambda $ is too small in order to proceed with the hard-collinear factorisation.  However,  it is useful to consider,  at least schematically, how the matrix element (\ref{meOoct})  can be  described in the limit $m_Q\to \infty$. 

 In this case, a proposal for the factorisation of this contribution  is shown in Fig.\ref{npQCD_graph}$\ b)$.  The corresponding  factorisation formula can be schematically  described  as 
\bea
 \left\langle VV \right\vert  C_h   \mathcal{O}_{oct}  \left\vert \chi_{cJ}\right\rangle  \sim  
 C_h \int d\omega_q \int d\omega_{\bar q} \ \Phi_\chi(\omega_q,\omega_{\bar q} ) 
\nonumber  \\
  \times \{J_n(\omega_q, x)*\phi_{3V}(x)\}\ \{ J_{\bar n}(\omega_{\bar q}, y)*\phi_{2V}(y)\} ,
\eea
where $\omega_i$ denotes the momentum fractions of the usoft quark fields, $J_{n,\bar n}$ are the appropriate jet functions, which are convoluted with the collinear LCDAs $\phi_{2,3V}$ (these convolutions  are  shown by the asterisks ). It is also supposed that the charmonium LDME is described by the LCDA $\Phi_\chi(\omega_1,\omega_2 )$, which is defined by the light-cone matrix element  
\bea
 \left\langle 0 \right\vert  \chi^{\dagger }T^a\sigma^\mu_{\top}\psi\  O_{\mu,a}(\bar q, q)   \left\vert \chi_{cJ}\right\rangle  
 \sim  \int d\omega_q \int d\omega_{\bar q}\  e^{i\omega_q (z_{q} n)+i\omega_{\bar q} (z_{\bar q}\bar n) } \Phi_\chi(\omega_q,\omega_{\bar q} ),
 \label{charmLCDA}
\eea
where the operator $O_{\mu,a}(\bar q, q) $ includes  usoft light quark $q(z_{q+})$ and antiquark $\bar q(z_{\bar q-})$ fields  describing the usoft charmonium cloud.    The structure of the operator in Eq.(\ref{charmLCDA}) indicates that the corresponding state consists of the octet-charm component and light degrees of freedom. Notice, that interactions of the usoft particles with the hard-collinear ones  make the NRQCD operator in Eq.(\ref{charmLCDA}) to be nonlocal.  In some sense, such charmonium LCDA reminds the  B-meson LCDAs $\varphi_{\pm}$, which describe  valence  heavy-light components of $b$-meson in the SCET factorisation framework.  

This very schematic picture is only  discussed  in order to stress  the relation of the matrix element  in  (\ref{meOoct}) with the colour-octet  component of the charmonium wave function and to illustrate  the common features of such configuration  with heavy-light  systems. 

The colour-octet matrix element can be parametrised  in the same way as the relevant amplitudes in Eqs.(\ref{def:A1}) and (\ref{def:A2}).  Therefore the final result for the amplitudes $A_{J}^{(LT)}$ is given by  the sum of  singlet and octet  contributions  
\begin{equation}
 A_{J}^{(LT)}=\sqrt{2M_\chi} \frac{f_{\chi} m_{V}}{m^3_{c}} \left(
\pi\alpha_{s}\right)  ^{2}\ \frac{2}{27}
\left\{ 
 \frac{f_{V}^{2}}{m_{c}^{2}} J^{(J)}_{V\Vert}+\frac{f_{V}^{\bot 2}}{m_{c}^{2}} J^{(J)}_{V\bot}
\right\}+a_{J}^{(LT)},
\label{ALTtot}
\end{equation}
where  $a_{J}^{(LT)}$ denotes the  contribution of the colour-octet matrix element.  We suppose that the endpoint singularities in the collinear integrals $J^{(J)}_{V\Vert,\bot}$ can be absorbed into UV-renormalisation of the octet amplitude $a_{J}^{(LT)}$. 

The power counting\footnote{ More details concerning the EFT power counting are given in Appendix~\ref{setup}  } for the colour-singlet contribution in Eq.(\ref{ALTtot}) gives  that it scales as  $A_{J}^{(LT)}\sim v^4 \lambda^3$.  Obviously,  the octet amplitudes $a_{J}^{(LT)}$ must have the same  behaviour.   The heavy quark component in the  colour-octet operator  scales as $v^3$ only. The additional factor of velocity $v$, which reproduces the balance appears from the interaction of the $c\bar c$ pair with the usoft gluon. The interaction of the  potential heavy quarks with  ultrasoft gluons is the subject of  potential NRQCD (pNRQCD).  The contributions of order $v$  in the effective Lagrangian are given by the dipole interactions,  which read $g_s\psi^\dagger(x) (-\boldsymbol{x}\cdot\boldsymbol{E}(t,\boldsymbol{0}))\psi(x)$ together with a similar term for the antiquark field \cite{Beneke:2008pi}. These interactions are  sensitive to the colour charge only and  conserve the spin and angular momentum of heavy quarks, which leads to the HQSS. The pNRQCD can not be used for the real calculations because of relatively  small  value of $m_c$. However this EFT gives us a possibility to better understand the power counting in the non-relativistic sector and the  dynamical origin of the HQSS.    

The colour-octet amplitudes $a_{J}^{(LT)}$ differ by the spin of quarkonium only and therefore they must satisfy  the symmetry relation up to higher order corrections in velocity $v$
\bea
a_{J=2}^{(LT)}=\lambda_{12} a_{J=1}^{(LT)}+\mathcal{O}(v^2),
\eea  
with a certain numerical coefficient $\lambda_{12}$.  This symmetry coefficient  can be directly calculated  in pNRQCD  in  the Coulomb limit $m_Q v^2\gg \Lambda$,  this analysis is carried out Sec.~\ref{Coulomb}.  On the other hand, $\lambda_{12}$ can also be derived from the colour-singlet contribution  by comparing the collinear integrals in the usoft limit.  

The simplest way to get $\lambda_{12}$  is to derive the relation between the physical amplitudes as described  above, see Eq.(\ref{A2A1T}). The divergent contributions must cancel in the symmetry breaking correction, therefore  using  Eq.(\ref{ALTcoll})  gives 
\bea
A_{2}^{(LT)}=\lambda_{12} A_{1}^{(LT)}+ \Delta A^{(LT)},
\label{HQSSALT}
\eea 
where 
\bea
 \Delta A^{(LT)}&=& \sqrt{2M_\chi} \frac{f_{\chi} m_{V}}{m^3_{c}} \left(
\pi\alpha_{s}\right)  ^{2}\ \frac{2}{27}
\\
&\times&
\left\{  
\frac{f_{V}^{2}}{m_{c}^{2}} 
 \left(
 J^{(2)}_{V\Vert}-\lambda_{12} J^{(1)}_{V\Vert}
 \right)
 +\frac{f_{V}^{\bot 2}}{m_{c}^{2}}
 \left ( 
 J^{(2)}_{V\bot}-\lambda_{12}J^{(1)}_{V\bot}
 \right)
\right\}.
\label{nnn}
\eea 
With the help of  Eqs.(\ref{J1s}) and (\ref{J2s}) one finds
\bea
 &&
 J^{(2)}_{V\Vert}-\lambda_{12} J^{(1)}_{V\Vert}
=\left( \frac 12 +\frac{\lambda_{12}}{\sqrt{2}} \right)
 J^{(s)}_{V\Vert}+
 \frac{\lambda_{12}}{\sqrt{2}} J^{(r)}_{V\Vert},
 \\
 && J^{(2)}_{V\bot}-\lambda_{12}J^{(1)}_{V\bot}
 =\left(1+ \lambda_{12}\sqrt{2} \right) J^{(s)}_{V\bot}.
 \eea
 The coefficients in front of divergent integrals  $J^{(s)}_{V\Vert,\bot}$ must vanish that gives
 \bea
 \lambda_{12}=-\frac{1}{\sqrt{2}}.
 \label{lam12}
 \eea
 Notice, that this  results holds for the two independent integrals simultaneously.   In Sec.~\ref{Coulomb}  we reproduce this result using pNRQCD calculation of the octet matrix element.   Therefore the final expression for the correction $\Delta A^{(LT)}$  in Eq.(\ref{HQSSALT})  reads
 \bea
 \Delta A^{(LT)}&=& \sqrt{2M_\chi} \frac{f_{\chi} m_{V}}{m^3_{c}} \left(
\pi\alpha_{s}\right)  ^{2}\ \frac{2}{27}\left(- \frac{1}{2}\right) J^{(r)}_{V\Vert},
\label{DALT}
\eea
 where the regular integral  $J^{(r)}_{V\Vert}$ is defined in Eq.(\ref{JrVVert}).  
 
 And finally, let's add a comment to the case $K^*\bar K^*$ channel. In this case the LCDAs are not symmetrical with respect to $x\leftrightarrow \bar x$ because of  $SU(3)$-symmetry breaking effects.  In addition, the twist-3 projections $M_3$ and  twist-3 LCDAs explicitly involve the contributions $\sim m_s/m_{K^{*}}$.  The expression  for the amplitude  in Eq.(\ref{ALTtot}) and relation (\ref{HQSSALT}) are obtained taking into account such terms. We find  that the contributions $\sim m_s$ occur in the final result only through the twist-3 LCDAs.  For simplicity,   such terms are not shown in the Eqs. (\ref{G3VIIu0})-(\ref{G3Vbot-x0}) describing  the endpoint  behaviour.  The inclusion of such terms complicates the formulas, but does not affect the main results.  For instance,   the LCDA  $G_{3K^*}^{\bot}$ in (\ref{G3VIIu0})  in the endpoint region is described as 
 \begin{align}
G_{3K^*}^{\bot}(x\rightarrow0) &=x [G_{3K^*}^{\bot}]'(0)+\ x \ln x \ \frac{f^\bot_{K^*}}{f_{K^*}}\frac{m_s}{m_{K^*}} \frac 12 [\phi^\perp_{2K^*}]'(0)+\mathcal{O}(x^2),
\label{G3KIIu0}
\end{align}
 where the first term in the {\it rhs} is given in Eq.(\ref{G3VIIx0}).  The endpoint behaviour of the correction in (\ref{G3KIIu0}) is not analytic $\sim x\ln x$, which gives the double logarithmic  contribution $\sim \ln^2\eta_{UV}/\delta_{IR}$ for the corresponding usoft  integrals.  It is interesting to note, that the similar  behaviour is also observed for the  amplitudes in $\chi_{cJ}\to K^*\bar K$ decays \cite{Kivel:2018rgd}, which are proportional to $SU(3)$ breaking contributions. However, in $K^*\bar K^*$ decays, the numerical effect from such  corrections is relatively small and  their detailed discussion is omitted  for simplicity.

\subsection{The hard kernels and HQSS relations for  amplitudes  $A_0^{TT}$ and $A_{2s}^{TT}$}

The calculation of these amplitudes involves the twist-3 and twist-4 LCDAs. The matrix elements of the valence twist-4 LCDAs $G_{4V}$  are defined as the matrix elements of the chiral-odd operator
\begin{equation}
 \langle V(k') \vert  \bar \chi_n (z_1) \Dsl{\bar{n}}\gamma_{\perp}^ \sigma \partial^\alpha_{\bot} \partial^{\beta}_\bot  \chi_n(z_2)  \vert 0\rangle,  \, z_i^2=0, 
 \label{Otw4}
\end{equation}
 see  details in Appendix~\ref{def:LCDA}.   The  analytical expression for the amplitudes is similar to one in Eq.(\ref{calc:AJLT}) but with the appropriate projections $M_i$ for the collinear matrix elements.  This calculation is also very similar to the previous ones, so we omit the technical details and present the result
 \bea
A_{J}^{(TT)}=\sqrt{2M_\chi} \frac{ f_{\chi}m_{V}^{2} }{m_{c}^{4}}\left(\pi\alpha_{s}\right)^{2} \frac{2}{27}\left\{
 \frac{f_{V}^{2}}{m_c^2} I_{33}^{(J)}+ \frac{ f_{V}^{\bot 2} }{m_c^2} I_{24}^{(J)}\right\}  ,
 \label{ATTcols}
\eea
The  colour-singlet contributions  are described by  $G_{3V}\otimes G_{3V}$ and  $\phi^\bot_{2V}\otimes G_{4V}$ configurations, which define the  collinear integrals $I_{33}$ and $I_{24}$, respectively.  Their explicit  expressions read ($D\equiv x y +\bar x \bar y$) 
\begin{align}
I_{33}^{(J)}  & =\int_0^1 dx \int_0^1 dy\ \frac{1}{x \bar x y \bar y D}
\left\{  \left(  G_{3V}^{\bot}(x)G_{3V}^{\bot}(y)+\tilde G_{3V}^{\bot}(x)\tilde G_{3V}^{\bot}(y)\right)  K_{J}(x,y) \right.
\\  &
\phantom {\int_0^1 dx \int_0^1 dy\ \frac{1}{x \bar x y \bar y D}}
\left. + \left( \tilde G_{3V}^{\bot}(x)G_{3V}^{\bot}(y)+G_{3V}^{\bot}(x)\tilde G_{3V}^{\bot}(y)\right)  \tilde{K}_{J}(x,y)\right\}  ,
\end{align}
with
\bea
K_{0}(x,y)&=&-\frac{1}{\sqrt{3}}\frac{1}{D}\left( {3}-\frac{2(x-\bar y)^{2}%
}{D}+(1-D)\frac{2D+(x-\bar y)^{2}}{4x\bar x y\bar y}\right) ,
\\
\tilde{K}_{0}(x,y)&=&-\frac{1}{\sqrt{3}}(x-\bar y)
\frac{(x-\bar y)^{2}+D-2}{4 x\bar x y\bar y D}.
\eea
\bea
K_{2s}(x,y)&=&-\frac{2(x-\bar y)^{2}}{D^{2}}-(1-D)
\frac{D-(x-\bar y)^{2}}{4x \bar x y\bar y D},
\\
\tilde{K}_{2s}(x,y)&=&\frac{(x-\bar y)\left(y-x \right)^{2}}{4x \bar x y \bar y D}.
\eea
\bea
I_{24}^{(0)}&=&-\frac{1}{\sqrt{3}}\frac1{16}\int_0^1 dx\ \frac{G^{(2)}_{4V}(x)}{\left(
x \bar x \right)  ^{2}}\int_0^1 dy \frac{\phi^{\bot}_{2V}(y_{i})}{y \bar y}
\frac{1}{D^{2}}\left(  3-x \bar y-y\bar x+8x\bar x\right)  ,
\\
I_{24}^{(2s)}&=&-\frac1{16}\int_0^1 dx \frac{G^{(2)}_{4V}(x)}{\left(  x\bar x\right) ^{2}
}\int_0^1 dy \frac{\phi^{\bot}_{2V}(y)}{y \bar y}\frac{1}{D^{2}}\left(
x \bar y+y \bar x-4x\bar x\right)  .
\eea
 For simplicity,  the corrections associated with the explicit $SU(3)$-breaking contributions  $\sim m_s/m_V$ are neglected in these formulas .  
 
 The convolution integrals $I_{24}^{(J)}$ and  $I_{33}^{(J)}$  are also singular in the endpoint regions where one of the gluons is soft  $x\sim \bar y\sim \lambda$ or  $\bar x\sim  y\sim \lambda$,  see Fig.\ref{pQCD_graphs}.  This can be seen in the same way as in the previous section.  The endpoint behaviour of the twist-3  LCDAs was already discussed,  using Eqs. (\ref{G3VIIu0})-(\ref{tG3VIIu1})  one finds \footnote{For simplicity, we assume that LCDAs  in Eqs.(\ref{GvGvx0})-(\ref{G4Vx0}) satisfy the symmetry $[\phi^{\Vert}_{2V}]'(0)=-[\phi^{\Vert}_{2V}]'(1)$, which does not hold for LCDAs of $K^*$ mesons, however the final results also valid  for $K^*\bar K^*$ case. }
\bea
\left.
 G^{\bot}_{3V}(x)G_{3V}^{\bot}(y)+\tilde G_{3V}^{\bot}(x)\tilde G_{3V}^{\bot}(y)\right |_{x\sim \bar y\sim \lambda}
 &=& -\frac{1}{4}[ G^{\bot}_{3V}]'(0)[\phi^{\Vert}_{2V}]'(0) x\bar y \left(  x+\bar y\right)+\mathcal{O}(\lambda^4),
 \label{GvGvx0}
 \\
\left.
 G^{\bot}_{3V}(x)\tilde{G}_{3V}^{\bot}(y)+\tilde{G}_{3V}^{\bot}(x) G_{3V}^{\bot}(y)\right |_{x\sim \bar y\sim \lambda}
 &=&-\frac{1}{4}[ G^{\bot}_{3V}]'(0)[\phi^{\Vert}_{2V}]' (0) x\bar y\left(  x-\bar y \right)  +\mathcal{O}(\lambda^4).
\\
G^{(2)}_{4V}(x \sim \lambda)=\frac
{1}{2}x^{2}[G^{(2)}_{4V}]^{\prime\prime}(0)+\mathcal{O}(\lambda^3).
 \label{G4Vx0}
\eea
 With the help of these formulas  one finds 
 \bea
\left[  I_{33}^{(0)}\right]  _{us(x\sim \bar y\sim \lambda)} 
&=&\frac{2}{\sqrt{3}}\left[  I_{33}^{(2s)}\right]  _{us(x\sim \bar y\sim \lambda)}
\label{I33usL1} \\
 &=&\frac{1}{\sqrt{3}}[ G^{\bot}_{3V}]'(0)[\phi^{\Vert}_{2V}]' (0) \frac12
\int_{\delta_{IR}}^{\eta_{UV}}dx \int_{\delta_{IR}}^{\eta_{UV}}d \bar y
 \frac{1}{\left(  x+\bar y\right)^{2}}.
 \label{I33usL2}
\eea
 \bea
\left[  I_{24}^{(0)}\right]  _{us(x\sim \bar y\sim \lambda)} 
&=&\frac{2}{\sqrt{3}}\left[  I_{24}^{(2s)}\right]  _{us(x\sim \bar y\sim \lambda)}
\label{I24usL1} \\
 &=&\frac{2}{\sqrt{3}}[G^{(2)}_{4V}]^{\prime\prime}(0)[\phi^{\bot}_{2V}]' (0)
 \frac 1{16}
\int_{\delta_{IR}}^{\eta_{UV}}dx \int_{\delta_{IR}}^{\eta_{UV}}d \bar y
 \frac{1}{\left(  x+\bar y\right)^{2}}.
\eea
 The similar result also holds for the symmetric domain $\bar x\sim  y\sim \lambda$. The obtained  structure of the IR-divergencies is the  same as in the previous section: the usoft integrals yields the simple logarithms. 
 This result differs from one of ref.\cite{Chen:2012ih} where the double logarithmic and even a power 
  divergencies are obtained. The occurence of the endpoint power divergence  in  ref.\cite{Chen:2012ih} indicates about the problem in the calculations: it can either be an error or an invalid operation on divergent collinear convolution integrals, see {\it e.g.} a discussion in ref.\cite{Beneke:2002bs}. 
  
  Because the usoft structure is  similar to one in the previous section, we follow the same  factorisation scheme  by introducing the appropriate colour-octet  matrix elements. This suggests
  \begin{equation}
 A_{J}^{(TT)}=\sqrt{2M_\chi} \frac{f_{\chi} m^2_{V}}{m^4_{c}} \left(
\pi\alpha_{s}\right)  ^{2}\ \frac{2}{27}
\left\{ 
 \frac{f_{V}^{2}}{m_{c}^{2}} I^{(J)}_{33}+\frac{f_{V}^{\bot 2}}{m_{c}^{2}} I^{(J)}_{24}
\right\}+a_{J}^{(TT)},
\label{ATTtot}
\end{equation}
where  $a_{J}^{(TT)}$ again denotes the  contributions  of the colour-octet matrix element.  Following the way  as in the previous section  and using results (\ref{I33usL1}) and (\ref{I24usL1}) for the usoft  integrals,   we can  conclude that the colour-octet contributions satisfy the HQSS relation 
\begin{equation}
a_{0}^{(TT)}=\frac{2}{\sqrt{3}}a_{2s}^{(TT)}+\mathcal{O}(v^2). 
\end{equation}
This result  yields  the desired relationship  between the decay amplitudes 
\begin{equation}
A_{0}^{(TT)}=\frac{2}{\sqrt{3}}A_{2s}^{(TT)}+\Delta A^{(TT)},
\label{A0A2TT}
\end{equation}
where  the  symmetry breaking corrections are described as 
\bea
\Delta A^{(TT)}= \sqrt{2M_\chi}\frac{f_{\chi}m_{V}^{2}}{m_{c}^{4}}
\left( \pi\alpha_{s}\right)^{2}\ \frac{2}{27} 
\left(
 \frac{f_V^2}{m_c^2} \Delta I_{33}
  + \frac{ f_{V}^{\bot2}}{m_c^2} \Delta I_{24}
  \right)  ,
 \label{HQSSATT}
\eea
with the well defined integrals ($D\equiv x y +\bar x \bar y$)
\bea
\Delta I_{24}&=&-\frac{1}{16}\frac{1}{\sqrt{3}}
\int_0^1 dx \frac{G^{(2)}_{4V}(x)}
{\left
( x \bar x\right)^{2}}
\int_0^1 dy \frac{\phi^{\bot}_{2V}(y)}{y \bar y}\frac{3D+16x \bar x}{D^{2}},
\\
\Delta I_{33}  & =&\int_0^1 dx \int_0^1 dy\ \frac{1}{x \bar x y \bar y D}
\left\{  \left(  G_{3V}^{\bot}(x)G_{3V}^{\bot}(y)+\tilde G_{3V}^{\bot}(x)\tilde G_{3V}^{\bot}(y)\right) \Delta K(x,y) \right.
\nonumber \\ && 
\phantom {\int_0^1 dx \int_0^1 dy\ \frac{1}{x \bar x y }}
\left. + \left( \tilde G_{3V}^{\bot}(x)G_{3V}^{\bot}(y)+G_{3V}^{\bot}(x)\tilde G_{3V}^{\bot}(y)\right)\Delta  \tilde{K}(x,y)\right\}  ,
\eea
with the kernels%
\bea
\Delta K(x,y)&=&\frac{\sqrt{3}}{D}\left(  -1+2\frac{(x-\bar y)^{2}}{D}
-\frac{1}{4}\left(  1-D\right)  \frac{(x-\bar y)^{2}}{x \bar x y  \bar y}\right)  ,
\\
\Delta\tilde{K}(x,y)&=&\frac{\sqrt{3}}{4}\frac{x-\bar y}{x \bar x y \bar y} 
\left(  
1-\frac{(x-\bar y)^{2}}{D }
\right)  .
\eea

\newpage
\section{Colour-octet matrix element in the Coulomb limit }
\label{Coulomb}

In this section we  calculate  the colour-octet matrix element defined in Eq.(\ref{meOoct}) in the Coulomb limit $m_Q\to \infty$ assuming that   $m_Q \gg m_Q v\gg m_Q v^2\gg \Lambda$.  Within this limit  quarkonium state can be considered  as a bound state created by the colour-singlet perturbative potential.  The ultrasoft modes become perturbative $p_{us}^2\gg \Lambda^2$ and  therefore  potential NRQCD  (pNRQCD)\cite{Pineda:1997bj, Pineda:1997ie,Brambilla:1999qa,Brambilla:1999xf, Brambilla:2004jw} can be used  in order to perform  calculations.  The main our task is to demonstrate that  octet amplitudes are restricted by HQSS and that their ultraviolet behaviour correctly reproduces the endpoint limit  of the colour-singlet  amplitude.  These results are universal and therefore can also be applied  for a realistic  case.  

Our calculation is very close to ones carried out in refs.\cite{Beneke:2008pi,Kivel:2018rgd} and many useful technical details can be found there.  We are not going  to perform a complex factorisation program as described in Fig.~\ref{npQCD_graph}$(b)$, instead,  we suppose as in ref.\cite{Beneke:2008pi} that  the final hadronic state is described by the collinear quark and antiquark where one constituent (associated with the ultrasoft  mode) has the  collinear fraction of order $v^2$. Such  collinear fraction is still larger than the nonperturbative  scale $\lambda\sim\Lambda/m_Q$  and therefore corresponding particle still can be associated with the collinear particle.  Formally, this allows one to describe the  hadronic matrix elements in terms of LCDAs even in the endpoint domain where the momentum fraction $x\sim v^2\gg \lambda$. This also implies that the  LCDAs  must be expanded  with respect to this  small fraction in order to avoid power corrections in small velocity $v$.  

Let us we consider  the calculation of the colour-octet contributions to the amplitudes $A_J^{(LT)}$. The  corresponding octet matrix elements  are defined as 
 \bea
  \left\langle V(k, e_V)V(k^{\prime},e'_V)\right\vert C_h \mathcal{O}_{oct}  \left\vert \chi_{QJ}
(P,\epsilon_{\chi})\right\rangle=i(2\pi)^{4}\delta^{(4)}(P-k-k^{\prime}) a_{J}
 \eea
 where definition of the operator  is given in Eq.(\ref{def:Ooct}). The scalar amplitudes are defined as in Eqs.(\ref{def:A1}) and (\ref{def:A2})
 \begin{align}
 a_{1}&=\left\{  i\varepsilon\lbrack\epsilon_{\chi},e_{V}^{\ast}]\left(e_{V}^{\ast\prime}\cdot k\right)  -i\varepsilon\lbrack\epsilon_{\chi}%
,e_{V}^{\ast\prime}]\left(  e_{V}^{\ast}\cdot k^{\prime}\right)  \right\} \frac{m_{V}}{M^2_{\chi}}a_{1}^{(LT)},
\label{def:a1LT}
\\
a_{2} &  = \epsilon_{\chi}^{\mu\nu}\left\{  k_{\mu}\left(  e_{V}^{\ast}\right)
_{\nu}^{\bot}\left(  e_{V}^{\ast\prime}\cdot k\right)  +k_{\mu}^{\prime
}\left(  e_{V}^{\prime\ast}\right)  _{\nu}^{\bot}\left(  e_{V}^{\ast}\cdot
k^{\prime}\right)  \right\}  \frac{4}{(k-k^{\prime})^{2}}\frac{m_{V}}{M_{\chi}}a_{2}^{(LT)}+\dots.
\label{def:a2LT}
\end{align}%
 
\begin{figure}[h!]
\centering
\includegraphics[width=4in]{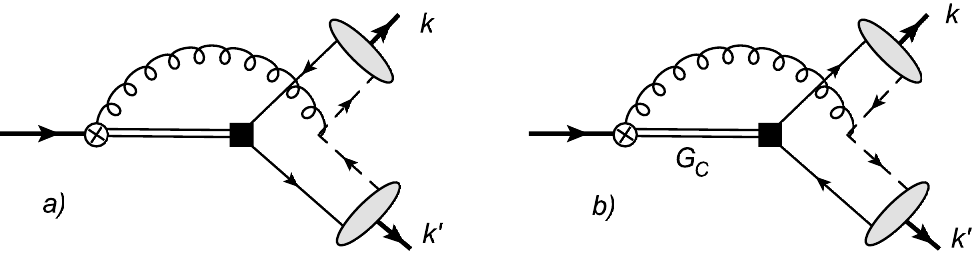}
\caption{
\label{pNRQCD_graphs}   Two diagrams, which describe the leading-order  pNRQCD  contribution to the colour-octet matrix element in the Coulomb limit. The crossed vertex denotes the dipole interaction (\ref{dipoleV}), the dashed lines correspond to  collinear particles with small fractions $\sim v^2$,  the double line denotes  the Coulomb Green function $G_c$ and the black square is associated with the octet operator vertex. 
}
\end{figure}

  The main  reorganisation in  the description of the usoft dynamics in this case  is associated with the heavy quark sector.  Integration over potential gluons gives pNRQCD \cite{Pineda:1997ie, Brambilla:1999qa,Brambilla:1999xf},  which  describes interactions of the bounded heavy  $Q \bar Q$ pair with the ultrasoft gluons.  The corresponding effective action is described in refs.\cite{Brambilla:1999xf, Brambilla:2004jw}.  The pNRQCD Lagarangian  includes  quark-antiquark colour-octet $\text{O}^a_{\beta\alpha}=[\chi^\dagger_\alpha T^a \psi_\beta]$ and colour-singlet $\text{S}_{\beta\alpha}=[\chi^\dagger_\alpha  \psi_\beta]$ effective fields  and  usoft gluon field.  The diagrams describing the colour-octet matrix element  are shown in Fig.~\ref{pNRQCD_graphs}. 
They  are generated by insertion of the  dipole interaction of order $v$, which in the notation of refs.~\cite{Brambilla:1999xf, Brambilla:2004jw}  reads
 \bea
 \mathcal{L}_{int}=\int d^3\boldsymbol{r}\ \text{Tr}[\text{O}^\dagger \text{\bf r}\cdot g_s\text{\bf E} \text{ S} +h.c. ]
 \label{dipoleV}
 \eea
 This vertex convert the  colour-singlet $Q\bar Q$ state into octet one with the usoft gluon. The occurring  colour-octet pair is  annihilated by the octet operator $\mathcal{O}_{oct}$.  The  Coulomb Green function $G_c$  is known  and its analytical expression is somewhat complicated and can be found in refs.~\cite{Schwinger:1964zzb, Beneke:2013jia}.
 
 Two diagrams in Fig.~\ref{pNRQCD_graphs} $(a,b)$ describe two different regions $x_1\sim y_2\sim v^2$ and $x_2\sim y_1\sim v^2$, respectively\footnote{ Remind, that  $x_1\equiv x, \ x_2\equiv 1-x$ and similar for the fractions $y_i$}.  The calculation of the both diagrams is very similar and therefore we only consider the diagram $(a)$ for simplicity. The analytic expression for this contribution can be written as 
\bea
a_J^ {(a)}
 &=&(-i)(-4\pi \alpha_s)\frac{2C_h}{27} \sqrt{2M_\chi}\sqrt{\frac{3N_c}{2\pi}} 
  \int \frac{d^{3}\bsd}{(2\pi)^{3}}~\tilde R_{21}(|\bsd|) 
  \int_{0}^{\eta}dx_{1}\int_{0}^{\eta}dy_{2}
  \nonumber \\  &&
\times  \frac{(-i)D_{q}^{\alpha\beta}}{(k_{1}+k_{2}^{\prime})^{2}} \frac{1}{4}\text{tr}\left[  \Lambda_{J}
(1-\Dsl\omega)\gamma_{\bot\beta}(1+\Dsl\omega)\right]  D^{Q}_{\alpha}(E,\bsd),
\label{octmeC}
\eea
where  all colour traces are calculated giving the factor 2/27. 
 This  expression includes the usoft gluon propagator 
\bea
\frac{(-i)}{(k_1+k'_2)^2}=\frac{(-i)}{2(kk')}\frac{1}{x_1y_2}.
\eea
The integration region over the usoft fractions  $x_1\sim y_2\sim v^2$  is restricted by UV-cut off $\eta$.  

The momentum space radial wave function of $P$-wave state reads
\bea
\tilde{R}_{21}(|\bsd|)=R_{21}^{\prime}(0)\frac{16\pi\gamma_{B}|\bsd|}{(\bsd^{2}+\gamma_{B}^{2}/4)^{3}},~\ \ \ \gamma_{B}=\frac{1}{2}%
m_{Q}\alpha_{s}C_{F} ,
\eea
where $R_{21}^{\prime}(0)$ is the derivative of the  position radial wave function at the origin. The angular integration yields
\bea
\int \frac{ d^{3}\bsd}{(2\pi)^3}~\tilde{R}_{21}(|\bsd|)|\bsd|=3 R'_{21}(0).
\label{def:N}
\eea

The factor $D_{q}^{\alpha\beta}$ in Eq.(\ref{octmeC}) describes  the trace associated with the collinear projectors and LCDAs. It can be computed in the same way as in Eq.(\ref{calc:AJLT}) but  now one must  also expand the obtained expression with respect to the small usoft fractions.  

The traces tr$\left[  \Lambda_{J}(1-\Dsl\omega)\gamma_{\perp\alpha}(1+\Dsl\omega)\right]$ describe the contractions of the  Dirac indices associated with the heavy quarks.   The projections onto $P$-wave states read
\bea
(1+\Dsl\omega)\Lambda_{J}  =-\frac{\Delta_{\top\mu}}{|\bsd|}\mathcal{P}_{J}^{\mu},  
\label{LJ}
\eea
where the projectors $\mathcal{P}_{J}^{\mu}$  are given in Eq.(\ref{def:PJ}). The transverse matrix $\gamma_\perp$ in these traces occurs from the colour-octet operator, see Eq.(\ref{def:Ooct}).

The function  $D^{Q}_{\beta}(E,\bsd)$  in Eq.(\ref{octmeC}) is described by the  interaction vertex (\ref{dipoleV})  and by octet Coulomb Green function $G_{c}$
\bea
D^{Q}_{\beta}(E,\Delta_{\top})=g_s (\omega_\beta k^\lambda_g-(g_{\top})_\beta^{\ \ \lambda}(\omega k_g))
 \frac{\partial}{\partial \bsd^\lambda}
 \int \frac{d^3\bsd'}{(2\pi)^3}  G_{c}(\bsd,\bsd'; E-(\omega k_g) ),
 \label{def:DQ}
\eea
where $k_g$  is  the outgoing ultrasoft  gluon momentum.  The expression for  the Green function is quite complicated,
 but we need only those terms, which give the UV-divergent integrals ($\sim \ln\eta$) in the diagrams  in Fig.~\ref{pNRQCD_graphs}.  It turns out that in this case  the Green function can be simplified
 \bea
G_{c}(\bsd,\bsd'; E_Q )=-\frac{(2\pi)^3\delta(\bsd-\bsd')}{E_Q-\bsd^2/m_Q}+\dots,
 \label{def:GC}
 \eea   
where $E_Q$ is the energy of heavy quark in the rest frame of $Q\bar Q$ pair, $\bsd$ denote the heavy quark relative momentum. The  dots denote the  contributions, which provide $UV$-regular expressions and therefore can be ignored for now. Substituting  $G_c$ from Eq.(\ref{def:GC}) into Eq.(\ref{def:DQ})  one finds
\bea
D^{Q}_\beta(E,\Delta_{\top})&\simeq&  g_s(\omega_\beta (\Delta_{\top}k_g)-\Delta_{\top\beta}(\omega k_g))\frac{2}{m_Q}\frac{1}{\left[  E-(\omega k_{g})-\bsd^2/m_Q+i\varepsilon\right]^{2}}
\\
&=&\Delta_{\top}^\lambda D^{Q}_{\beta\lambda},
\label{DQ}
\eea
where $E$  denotes  the binding energy $E=M_\chi-2m_Q$ for $P$-wave triplet.\footnote{ To our accuracy the value of $E$ is degenerated and therefore does not depend on the total angular momentum $J$ of the perturbative state $\chi_{QJ}$. } Using Eqs.(\ref{LJ}) and (\ref{DQ}) and rotation invariance ($ \Delta^\lambda_{\top}\Delta^\mu_{\top}\to -g_\top^{\mu\lambda}\bsd^2/3 $) one can rewrite the integrand
\bea
 \frac{1}{4}\text{tr}\left[  \Lambda_{J}
(1-\Dsl\omega)\gamma_{\bot\alpha}(1+\Dsl\omega)\right]  D^{Q}_{\beta}(E,\bsd) 
 = \frac{|\bsd|}{3} D^{Q}_{\beta\mu}
 \frac{1}{4}\text{tr}\left[ \mathcal{P}_{J}^{\mu}\gamma_{\bot\alpha}\right]  .
\eea
 This gives
 \bea
 a^{(a)}_J &=&4\pi \alpha_s \frac{2C_h}{27} \frac{1}{2(kk')} \sqrt{2M_\chi}\sqrt{\frac{3N_c}{2\pi}} 
  \int \frac{d^{3}\bsd}{(2\pi)^{3}}~\tilde R_{21}(|\bsd|) \frac{|\bsd|}{3}
 \nonumber \\  && \times 
  \int_{0}^{\eta}\frac{dx_{1}}{x_1}\int_{0}^{\eta}\frac{dy_{2}}{y_2}
 \  D_{q}^{\alpha\beta}\  D^{Q}_{\beta\mu}\ \frac{1}{4}\text{tr}\left[ \mathcal{P}_{J}^{\mu}\gamma_{\bot\alpha}\right].
\label{aJ}
\eea
 In order to get expressions for the scalar amplitudes $a^{(LT)}_i$, it is enough to consider the contribution to the  factor $D_{q}^{\alpha\beta}$  with the projections of twist-3 for the state $V(k,e)$ and twist-2 for $V(k',e')$ only. This  yields 
\bea
\left. D_{q}^{\alpha\beta}\right |_{x_1\sim y_2\sim v^2} &=&  \text{Tr}[\gamma^\alpha \hat M_3(x_1, k) \gamma^\beta \hat M_2(y_2,k')] 
\nonumber \\
&=& y_2 \frac{1}{4}\frac{ \left(  ke^{\prime\ast}\right)  }{(kk^{\prime})}\
 \left(  f_{V}m_{V}\right)^{2} F_{\Vert}\left(  e_{\bot}^{\ast}\right)  ^{\alpha}k^{\prime\beta}
\nonumber \\ 
&-&y_2 \frac{1}{4}\frac{ \left(  e^{\ast}k^{\prime}\right) }
{(kk^{\prime})}( f_{V}^{\bot}m_{V})^2 \ F_{\bot}\left(  e_{\bot}^{\prime\ast}\right)
^{\alpha} k^{\prime\beta}+\mathcal{O}(v^4),
\eea
with
\begin{align}
F_{\Vert} &  =\
\left\{   [G_{3V}^{\bot}]'(0)-[\tilde G_{3V}^{\bot}]'(0) \right\}   
[\phi^\Vert_{2V}]'(1),
\\
F_{\bot} &  =
 [G_{3V}^{\Vert}]'(0)[\phi^{\bot}_{2V}(1)]'.
\end{align}
Substituting this into Eq.(\ref{aJ})  one finds
\bea
a^{(a)}_J  & =&-C_h (4\pi\alpha_s) \frac{2}{27}\sqrt{M_\chi}\frac{1}{2(kk')}\frac{1}{m^2_Q}
\text{tr}\left[  \mathcal{P}_{J}^{\mu}\gamma_{\bot}^{\alpha}\right] k^{\prime\mu}
\nonumber \\
&\times& \frac14 \left[ 
 \frac{ \left(  ke^{\prime\ast}\right)  }{(kk^{\prime})}\
 \left(  f_{V}m_{V}\right)^{2}  F_{\Vert} \left(  e_{\bot}^{\ast}\right)^{\alpha}
  -\frac{ \left(  e^{\ast}k^{\prime}\right) }
{(kk^{\prime})}( f_{V}^{\bot}m_{V})^2F_{\bot} \left( e_{\bot}^{\prime\ast}\right)^{\alpha}
 \right] J_{oct},
 \eea
 where  the universal integral  reads
\bea
 J_{oct}&=&\sqrt{\frac{3N_c}{2\pi}}\int \frac{d^3\bsd}{(2\pi)^3}~R_{21}(|\bsd|)
 ~\frac{\left\vert \bsd \right\vert}{3}
 \nonumber \\ &\times&  \int_{0}^{\eta}dx_{1}\int_{0}^{\eta}dy_{2} 
 \frac{m_Q^2}{\left[  E-m_Q(x_{1}+y_{2})-\bsd^{2}/m_Q+i\varepsilon
\right]^{2}} .
\label{Joct}
\eea
Performing further calculations and matching with Eqs.(\ref{def:a1LT})-(\ref{def:a2LT})  one finds
\bea
a^{(LT,a)}_{1}&=&\sqrt{2M_\chi} \frac{ m_{V}}{m^3_{Q}} 
\pi^2\alpha_{s}(\mu_h)\alpha_{s}(\mu_{us}) \frac{2}{27} 
 (-\sqrt{2} )\left( 
\frac{1}{2}\frac{ f^2_V }{m_Q^2} F_{\Vert}  +\frac{ (f^\bot_V)^2}{m_Q^2}  F_{\bot}
\right) J_{oct},
\label{a1a}
\\
a^{(LT,a)}_{2}&=&\sqrt{2M_\chi} \frac{ m_{V}}{m^3_{Q}} 
\pi^2\alpha_{s}(\mu_h)\alpha_{s}(\mu_{us}) \frac{2}{27} \left( 
\frac{1}{2}\frac{ f^2_V }{m_Q^2} F_{\Vert}  + \frac{ (f^\bot_V)^2}{m_Q^2}  F_{\bot}
\right) J_{oct}.
\label{a2a}
\eea
From this expressions it follows  that 
\bea
a^{(LT,a)}_{2}=-\frac{1}{\sqrt{2}}a^{(LT,a)}_{1},
\eea
which is in agreement with the result (\ref{lam12}). The same result also valid for the contributions of the second diagram in Fig.~\ref{pNRQCD_graphs}.

In order to demonstrate the overlap of the IR-divergencies in Eqs.(\ref{JsVbotUS}) and (\ref{JsVVertUS1}) with UV-divergencies in Eqs.(\ref{a1a}) and (\ref{a2a}) let us consider the integral in Eq.(\ref{Joct}).  Taking the formal limit $\eta\to\infty$ and assuming that $m_Q x_1\sim m_Q y_2\gg E\sim \bsd^2/m_Q\sim v^2$ one can neglect small terms in the denominator of the integrand in (\ref{Joct})  that gives
\bea
\left.J_{oct}\right|_{UV}&\simeq &\sqrt{\frac{3N_c}{2\pi}}\int \frac{d^3\bsd}{(2\pi)^3}~R_{21}(|\bsd|)
 ~\frac{\left\vert \bsd \right\vert}{3}
 \int_{0}^{\infty}dx_{1}\int_{0}^{\infty}dy_{2} \frac{1}{\left( x_{1}+y_{2}\right)^{2}} 
 \\ 
 &=&\sqrt{\frac{3N_c}{2\pi}}R'_{21}(0)  \int_{0}^{\infty}dx_{1}  \int_{0}^{\infty}dy_{2}\ \frac{1}{\left( x_{1}+y_{2}\right)^{2}}
,
\label{Joct_UV}
\eea
where  we used Eq.(\ref{def:N}). Substituting this result into Eqs.(\ref{a1a}) and (\ref{a2a}) one can easily observe that obtained expression correctly reproduces the corresponding IR-asymptotic  (\ref{JsVbotUS}) and (\ref{JsVVertUS1})  of the   colour-singlet  contribution in Eq.(\ref{ALTcoll}). 

Let us also note, that  the integrals  $d^3\bsd $ in Eqs.(\ref{a1a}) and  (\ref{a2a}) are well defined. The denominator of the Coulomb Green function  has the pole at $\bsd^2/(2m_Q)=E-(x_1+y_2)m_Q$, which leads to the nontrivial  imaginary phase. The obtained in this way imaginary part does not have UV-divergencies at $\eta\to \infty$ as it must be.  This observation illustrates an interesting mechanism leading to the nonperturbative phases in QCD.

\section{Phenomenology}
\label{phenom}
The helicity amplitudes $F^{(J)}_{\lambda,\lambda'}$, which describe  $\chi_{cJ}\to V(\lambda) V(\lambda')$ decays are defined, see {\it e.g.} refs.\cite{Huang:2021kfm, Chen:2013gka}
\bea
F^{(J)}_{\lambda,\lambda'}=\sum_i \rho_J(i) A_J(i,\lambda,\lambda'),
\eea
where spin-density matrix $\rho_J$  describing $\chi_{cJ}$ production read 
\begin{align}
\rho_0=1, \ \ \rho_{1}(i)=N_1 \text{ diag}\{1,2,1\},\ i=(-1,0,1), 
\\
\rho_{2}(i)=N_2\text{ diag}\{ 2,1,\frac{2}{3},1,2 \},\  i=(-2,-1,0,+1,+2).
\end{align}
The normalisations $N_i$  depend on the amplitude $\psi(2S)\to\chi_{cJ}+\gamma$, see more details in  ref.\cite{Chen:2020pia}.  These normalisations cancel in the observed polarisation parameters and therefore their explicit values are not important.  

In order to find expressions of $F^{(J)}_{\lambda,\lambda'}$ in terms of  scalar amplitudes $A_J^{(\dots)}$ we consider the rest frame of the parent charmonium $\chi_{cJ}$ and define the momenta as   
\bea
P=(M_\chi,\vec 0),  k=\frac12 M_\chi(1,0,0,\beta_V), k'=\frac12 M_\chi(1,0,0,-\beta_V).
\eea  
The polarisation vectors for different particles cab be defined as 
\bea
\epsilon^\mu_\chi(0)&=&(0,0,0,1),\ \ \epsilon^\mu_\chi(\pm)=\mp\frac{1}{\sqrt{2}}(0,1,\pm i,0),
\\
 \epsilon^{\mu\nu}_\chi(+2)&=&\epsilon^\mu_\chi(+)\epsilon^\mu_\chi(+), \ \   
 \epsilon^{\mu\nu}_\chi(+1)=\frac{1}{\sqrt{2}} 
 \left( 
 \epsilon^\mu_\chi(+)\epsilon^\mu_\chi(0)+\epsilon^\mu_\chi(0)\epsilon^\mu_\chi(+) 
 \right),
 \\
 \epsilon^{\mu\nu}_\chi(0) &=& \frac{1}{\sqrt{6}} 
  \left( 
 \epsilon^\mu_\chi(+)\epsilon^\mu_\chi(-)+\epsilon^\mu_\chi(-)\epsilon^\mu_\chi(+)+2 \epsilon^\mu_\chi(0)\epsilon^\mu_\chi(0)
 \right).
\\
e^\mu_V(\pm)&=& e^{\prime \mu}_V(\mp)=\pm\frac{1}{\sqrt{2}}(0,1,\pm i,0).
\eea
A simple calculation yields \begin{equation}
x=\frac{\left\vert F_{1,1}^{(0)}\right\vert }{\left\vert F_{0,0}%
^{(0)}\right\vert }=\frac{\left\vert A_{0}^{(TT)}\right\vert }{\left\vert
A_{0}^{(LL)}\right\vert }, \ \ 
y_{0}=\frac{|F_{1,-1}^{(2)}|}{|F_{00}^{(2)}|}=3\sqrt{2}\frac{\left\vert
A_{2t}^{(TT)}\right\vert }{\left\vert A_{2}^{(LL)}\right\vert }.
\label{defxy0}
\end{equation}%
\begin{equation}
y_{1}=\frac{\left\vert F_{0,1}^{(2)}\right\vert }{|F_{00}^{(2)}|}=\frac{3}%
{2\sqrt{2}}\frac{\left\vert A_{2}^{(LT)}\right\vert }{\left\vert
A_{2}^{(LL)}\right\vert }, \ \
y_{2}=\frac{\left\vert F_{1,1}^{(2)}\right\vert }{|F_{00}^{(2)}|}%
=\frac{\left\vert A_{2s}^{(TT)}\right\vert }{\left\vert A_{2}^{(LL)}%
\right\vert }.
\end{equation}
From  Eq.(\ref{hierarchy}) one finds 
\bea
y_0\sim \mathcal{O}(1), \  y_1\sim \mathcal{O}(\Lambda/m_c),  \  x\sim y_2\sim \mathcal{O}(\Lambda^2/m^2_c),
\eea
which implies
\bea
y_0 \gg y_1 \gg   x\sim y_2.
\label{yx_hier}
\eea

  The available experimental results  read \cite{BESIII:2023zcs}
\bea
x&=&0.30\pm 0.03,
\label{data:x}
\\
y_0&=&1.45\pm 0.10 ,  \  y_1=1.265\pm0.054, \  y_2=0.81\pm 0.05,
\label{data:yi}
\eea
where  only the statistical uncertainties are shown. Taking the conservative estimate $\Lambda/m_c\sim 0.3$, one finds that expected  hierarchy (\ref{yx_hier}) is violated. In particular, the value of $y_0$ is almost the same order as $y_1$, even despite the large numerical coefficient $3\sqrt{2}$ in the definition (\ref{defxy0}).  The value of $y_2$ also looks much larger than might be expected.  These observation indicates about the possible scaling violation effects in these decays. 

At the same time the  value of the parameter  $x$ in (\ref{data:x}) is still small  indicating that  the leading amplitude $A_0^{(LL)}$ dominates over the suppressed one $A_{0}^{(TT)}$ as expected.  A slightly larger value of $x \simeq 3 \Lambda^2/m_c^2$ than might be expected can be understood as the influence of the colour-octet contribution.  The  values of the partial widths can be easily found and reads
\bea
\Gamma_{0}^{(LL)} &=&\frac{\Gamma[\chi_{c0}\to \phi\phi]}{1+2x^{2}}=0.85\ \Gamma[\chi_{c0}\to \phi\phi]=7.55\text{ KeV},
\label{G0LLdata}
\\ 
\Gamma_0^{(TT)}&=&\frac{2x^{2}\Gamma[\chi_{c0}\to \phi\phi]}{2x^{2}%
+1}=0.15\ \Gamma[\chi_{c0}\to \phi\phi]=1.35 \text{ KeV}.
\label{G0TTnum}
\eea
where we used the value $Br[\chi_{c0}\to \phi\phi]=8.48\times 10^{-4}$ from ref. \cite{BESIII:2023zcs} and the total width 
$\Gamma[\chi_{c0}]=10.5\ $MeV \cite{Workman:2022ynf}.

On the other hand $\Gamma_{0}^{(LL)}$ can be estimated using result for the amplitude $A_0^{(LL)}$ in Eq.(\ref{Ai**}).  In order to get numerical estimates we use the following  input for different parameters. For the charmonium matrix element in (\ref{def:Rp21}) we use the value, which follows from a Buchmüller- Tye potential model in ref.~\cite{Eichten:1995ch}
\bea
\frac{1}{3}| \left\langle 0\right\vert \chi\dag\left(-\frac{i}{2}\overleftrightarrow{\boldsymbol{D}}\cdot\boldsymbol{\sigma }\right)\psi
\left\vert \chi_{c0}\right\rangle |^2 \approx 0.107 \times 10^{-2} \text{ GeV}^5. 
\eea
The averaged charmonium mass is fixed as $M_\chi=3.5\ $GeV.  For quark mass $m_c$  there is well known estimate  $m_c=1.4\pm 0.2\ $GeV.  The errors  in this expression  give large numerical effect  because  the expressions in (\ref{Ai**}) are proportional to $1/m_c^4$.  For simplicity, we  take  the value of $m_c=1.5\ $GeV. Various LCDA models are described in Appendix~\ref{def:LCDA}.  For the QCD running coupling we fix  the value $\alpha_{s}=0.25$.
 Then we find 
\begin{equation}
\Gamma_{0}^{(LL)}=\left(  8.2\pm2.7 \right)  \text{ KeV} ,
\end{equation}
 where the error indicates uncertainties from the  parameters of the LCDAs.  This value is in the good agreement with one obtained from the data (\ref{G0LLdata}) that supports our assumption about the reliable value of $x$.  
 
In order to better understand the different values of ratios $y_i$  let us estimate these parameters  using the theoretical  results for the different amplitudes  together with the data for $x$ and widths $\Gamma[\chi_{c0}\to \phi\phi]$ and $\Gamma[\chi_{c1}\to \phi\phi]$. 

First,  consider the ratio $y_0$. This parameter can be obtained from the results for the leading-order amplitudes 
$A_{2t}^{(TT)}$ and $A_{2}^{(LL)}$ in Eqs.(\ref{Ai**}),(\ref{J2LL}) and (\ref{J2tTT}). The  analytical  expression for the corresponding ratio reads
\begin{equation}
y_{0}^{\text{th}}=3\sqrt{2}\frac{\left\vert A_{2t}^{(TT)}\right\vert
}{\left\vert A_{2}^{(LL)}\right\vert }=3\sqrt{2}~\frac{f_{V}^{\bot2}}%
{f_{V}^{2}}\frac{\left\vert J_{2t}^{(TT)}\right\vert }{\left\vert J_{2}%
^{(LL)}\right\vert }=3\sqrt{2}\times\left(  0.872\pm0.21\right)
=3.70\pm0.91.
\end{equation}
The error in this formula is associated with the ambiguities of the hadronic parameters in LCDAs. One sees that the theory at this order predicts the ratio of the amplitudes $\vert A_{2t}^{(TT)}\vert/\vert A_{2}^{(LL)}\vert\sim 1$ and the large value of $y_{0}^{\text{th}}$ is closely associated with the large numerical coefficient $3\sqrt{2}$.  This gives  the value $y_{0}^{\text{th}}$, which  is about factor two larger than experimental one, see Eq.(\ref{data:yi}).  

In order to estimate $y_1$ we use the branching ratio $Br[\chi_{c1}\to \phi\phi]=4.48\times 10^{-4}$ from ref.\cite{BESIII:2023zcs}  and the total width $\Gamma[\chi_{c1}]=0.88\ $MeV \cite{Workman:2022ynf}. This allows one  to get  $|A_1^{(LT)}|$
\begin{equation}
\left\vert A_{1}^{(LT)}\right\vert =\sqrt{\Gamma\lbrack\chi_{c1}%
\rightarrow\phi\phi]\left[  \frac{1}{2}\frac{1}{16\pi}\frac{\beta_{\phi}}%
{M_{\chi}}\frac{1}{3}\right]  ^{-1}}\simeq23.79\ \text{MeV}.
\end{equation}
Then  the relationship (\ref{HQSSALT}) allows one  to estimate $|A_2^{(LT)}|$
\begin{equation}
\left\vert A_{2}^{(LT)}\right\vert =\left\vert A_{1}^{(LT)}\right\vert
\left\vert -\frac{\sqrt{2}}{2}+e^{-i\phi_1}\frac{\Delta A^{(LT)}}{\left\vert
A_{1}^{(LT)}\right\vert }\right\vert ,
\end{equation}
where $\phi_1$ is the unknown phase of the amplitude $A_{1}^{(LT)}$.  The value  $\Delta A^{(LT)}$  can be estimated from (\ref{DALT}) using the same hadronic parameters as for  $\Gamma_{0}^{(LL)}$. This gives 
\begin{equation}
\left\vert A_{2}^{(LT)}\right\vert \simeq\left\vert A_{1}^{(LT)}\right\vert
\left\vert -0.707~-0.573~e^{-i\phi_1}\right\vert .
\end{equation}
The  value of  $A_{2}^{(LL)}$ can also be estimated from Eq.(\ref{Ai**}) and this gives 
\begin{align}
y_{1}^{\text{th}}  & \simeq\frac{3}{2}\frac{1}{\sqrt{2}}\frac{\left\vert
A_{1}^{(LT)}\right\vert }{\left\vert A_{2}^{(LL)}\right\vert }\left\vert
-0.707~-0.573~e^{-i\phi_1}\right\vert =0.523\left\vert -0.707~-0.438~e^{-i\phi_1
}\right\vert .
\end{align}
Taking  $\phi_1=0$ in order to get the maximal estimate  for $y_{1}^{\text{th}}$ one finds
\bea
y_{1}^{\text{th}}  \simeq 0.73.
\eea
This value is approximately two times lower than $y_1$ in (\ref{data:yi}).  At the same time $y_{1}^{\text{th}}\simeq 2.5 \Lambda/m_c$ (remind, that above we accepted  $\Lambda/m_c\simeq 0.3$).

The estimate of $y_2$ is based on the relationship  (\ref{A0A2TT}).  The value of $|A_0^{TT}|$ can be obtained using formulas  (\ref{Gam1}) and  (\ref{G0TTnum})
\begin{equation}
\left\vert A_{0}^{(TT)}\right\vert =\sqrt{0.15\Gamma\lbrack\chi_{c0}%
\rightarrow\phi\phi]\left[  \frac{1}{2}\frac{1}{16\pi}\frac{\beta_{V}}%
{M_{\chi}}2\right]  ^{-1}}\simeq5.375\ \text{MeV}.
\end{equation}
Then using (\ref{HQSSATT}) and (\ref{data:x}) one finds 
\begin{equation}
\left\vert A_{2s}^{(TT)}\right\vert =\frac{\sqrt{3}}{2}\left\vert A_{0}^{(TT)}
\right\vert \left\vert 1-\frac{\Delta A^{(TT)}}{\left\vert A_{0}%
^{(TT)}\right\vert }e^{-i\phi_2}\right\vert
= 4.655\left\vert 1+2.150e^{-i\phi_2 }\right\vert ,
\end{equation}
where $\phi_2$ is unknown phase of the amplitude $A_{0}^{(TT)}$. 
Using theoretical estimate for $A_2^{(LL)}$ and  taking  $\phi_2=0$ in order to get maximal numerical  value  one finds
\begin{equation}
 y_{2}^{\text{th}}\ \simeq 0.33.
\end{equation}
This result is close in value with $x$ in Eq.(\ref{data:x}) and is about factor  $2-3$ lower than the experimental result in (\ref{data:yi}).  The obtained theoretical estimates satisfy the expected hierarchy 
\bea
\frac{y_{1}^{\text{th}} }{ y_{0}^{\text{th}}}\simeq 0.20 \sim \mathcal{O}(\Lambda/m_c),
\ \frac{y_{2}^{\text{th}} }{ y_{0}^{\text{th}}}\simeq 0.09\sim  \mathcal{O}(\Lambda^2/m^2_c).
\eea

The obtained values of  $y_{i}^{\text{th}}$ allows to estimate the total width $\chi_{c2}$
\begin{equation}
\Gamma^{\text{th}}\left[  \chi_{c2}\rightarrow\phi\phi\right]  =\Gamma_{2}^{(LL)}
\left(  1+\frac{2}{3}(y_{0}^{\text{th}})^{2}+\frac{8}{3}(y_{1}^{\text{th}})^{2}+2 ( y_{2}^{\text{th}})^2\right)
=  5.3\pm 2.2  \ \text{KeV},
\end{equation}
where the error  gives the uncertainty due to hadronic parameters in mesonic LCDAs. 
This estimate has to be compared with the experimental value 
$\Gamma\left[ \chi_{c2}\rightarrow\phi\phi\right]=2.12\ $KeV obtained from the data for  branching fraction in 
ref.\cite{BESIII:2023zcs}.   The  theoretical estimate  is  larger  but  have large theoretical uncertainty  and  for the leading-order accuracy can be accepted as a reasonable  result.  

However,  the obtained  values $y^{(\text{th})}_i$ predict the different partial widths compared to 
what can be obtained from data. The partial widths can be easily obtained from 
\bea
{\Gamma_{2}^{(LL)}}&=& {\Gamma[\chi_{c2}\to \phi\phi]} \left(1+\frac{2}{3}y_{0}^{2}+\frac{8}{3}y_{1}^{2}+2y_{2}^{2}\right)^{-1},
\\
\Gamma_{2t}^{(TT)}& =& \frac{2}{3}y_{0}^{2}\, \Gamma_{2}^{(LL)},
\ \
\Gamma_{2}^{(LT)}=\frac{8}{3}y_{1}^{2}\, \Gamma_{2}^{(LL)},
\ \
\Gamma_{2s}^{(TT)}=2y_{2}^{2}\, \Gamma_{2}^{(LL)}.
\eea
Numerical results for their values  are presented in Table.~\ref{tab2}. 
\begin{table}[h!]
\centering
 \caption{ Numerical  results for theoretical estimates of different  partial widths in comparison with the experimental values obtained using the data from ref.\cite{BESIII:2023zcs}  } 
 \label{tab2}
\begin{tabular}[c]{|c|c|c|}\hline
  & theory & experiment  \\\hline
 $ \Gamma_{2}^{(LL)}/\Gamma[  \chi_{c2}\rightarrow\phi\phi] $& 0.08  & 0.13 \\\hline
$\Gamma_{2t}^{(TT)}/\Gamma[  \chi_{c2}\rightarrow\phi\phi] $ & 0.78  &0.17  \\\hline
$\Gamma_{2}^{(LT)} /\Gamma[  \chi_{c2}\rightarrow\phi\phi] $  &0.12  &0.54  \\\hline
$\Gamma_{2s}^{(TT)} /\Gamma[  \chi_{c2}\rightarrow\phi\phi]$ &0.02  &0.16  \\\hline
\end{tabular}
\end{table}
One finds  that the  theoretical consideration predicts that  the $\chi_{c2}$ width is dominated by $\Gamma_{2t}^{(TT)}\sim
78\%$ and other contributions are relatively small. On the other hand, the data suggest that the largest contribution is provided by  
$\Gamma_{2}^{(LT)}\sim50\%$, which is  suppressed by $\Lambda^2/m_{c}^{2}$ in comparison with $\Gamma
_{2}^{(LL)}$ and $\Gamma_{2t}^{(TT)}$.  Moreover, the measured  value  $\Gamma_{2s}^{(TT)}$ is of the same order as the leading-power contributions $\Gamma_{2t}^{(TT)}$ and $\Gamma_{2}^{(LL)}$ 
while this contribution must be  suppressed as $\Lambda^{4}/m_{c}^{4}$.  

 These observations allow us to conclude that available data on  $\chi_{c2}\to \phi\phi$ decay indicate a strong deviation from  expected EFT description. At the same time,  the data on  $\chi_{c0,1}\to \phi\phi$  do not show a strong violation of the amplitude hierarchy predicted by the $1/m_c$ expansion.  Therefore  the  underlying dynamics describing  decays $\chi_{c0,1}$ and $\chi_{c2}$ can be very different.  At present, the possible dynamic nature of this effect is not clear to us.

\section{Summary}
\label{sum}

We have studied  the decays of $P$-wave charmonia into two vector mesons $\chi_{cJ}\to VV$.  
The covariant definitions of the amplitudes and analytical expressions for the  decay widths have been developed.   We used the QCD effective field theory approach in order to study the factorisation for various amplitudes. It is shown that the helicity suppressed amplitudes are described by the sum of the colour-singlet and colour-octet contributions. The endpoint divergencies arising  in the colour-singlet contributions in these amplitudes  can be absorbed by the renormalisation of the colour-octet contribution.  The latter is defined in terms of a SCET-I matrix element  and can be  associated  with  the  colour-octet component of the charmonium wave function in the limit $m_c\to \infty$.  

The  colour-octet  contribution depends  on the   intermediate hard-collinear scale $\sim m_c\Lambda$, which is relatively small for the realistic  value of the charm mass.  Therefore  the  SCET-I matrix elements  are considered in this work as non-perturbative  quantities.  It is shown these contributions satisfy certain relations determined by  the heavy quark spin symmetry up to higher order corrections in small velocity $v$.   Using this information and data  for $\chi_{c0}\to \phi\phi$ and $\chi_{c1}\to \phi\phi$ we obtained estimates for the polarisation parameters of  $\chi_{c2}\to \phi\phi$.  

Comparison of these estimates with the available data  shows a strong discrepancy.  The data  suggest a very large values  of the subleading amplitudes $F^{(2)}_{1,0}$  and $F^{(2)}_{1,1}$, which contradicts  the power counting in $\Lambda/m_c$.  This perhaps indicates dynamics that produce large numerical effects beyond the leading power approximation. Further research is needed to understand this phenomenon. In this situation, it would be very interesting to study whether the same problem arises for other vector meson decay channels.

\appendix
\section{Set up, useful notation and power counting}
\label{setup}
In this article  we use the frame where  heavy meson is at rest and the $z$-axis is
chosen along the momenta of  outgoing particles
\begin{equation}
P=M_\chi(1,\vec{0})=M_\chi\omega ,  \label{def:w}
\end{equation}%
where $M_\chi$ is charmonium mass  and $\omega $ denotes charmonium  four-velocity. 
Any four-vector $p$,   which is orthogonal  to velocity $\omega$ is denoted with the subscript  $\top$: $\omega \cdot p_{\top} =0$ where 
\bea
p_{\top}^\nu=p_\mu g_\top^{\mu\nu},  \  g_\top^{\mu\nu}=g^{\mu\nu}-\omega^\mu\omega^\nu.
\eea

The  non-relativistic sector involve the particles with the following momenta
\bea
\text{hard}&:&  p_0\sim \vec p \sim m_c ,
\\
\text{soft }&:&  p_0\sim \vec p \sim m_c v,
\\
\text{potential }&:&  p_0\sim  m_c v^2,\  \vec p \sim m_c v,
\\
\text{ultrasoft (usoft) }&:&  p_0\sim \vec p \sim m_c v^2.
\eea
For description of the NRQCD matrix elements  we use the standard notations from ref.\cite{Bodwin:1994jh}.  The details about pNRQCD can be found, for instance,  in refs.\cite{Brambilla:1999xf,Brambilla:2004jw}.  

The intermediate EFT SCET-I  describes the interaction of  usoft and hard-collinear particles. To describe this EFT  we use hybrid formulation and notations from ref.\cite{Beneke:2002ph}.  

Let us introduce  two auxiliary light-like vectors $n$ and $\bar n$, satisfying $n^2=\nb^2= 0$,  $(n\cdot \nb)= 2$.  Then the hard-collinear momentum  can be represented as 
\bea
p_{hc}^\mu=(p_{hc}\cdot n) \frac{\nb^\mu}{2}+ (p_{hc}\cdot \nb) \frac{n^\mu}{2}+p_{{hc}\bot} ^\mu,
\eea
where  the transverse component is given by 
\bea
p_{hc\bot} ^\mu=(p_{hc})_\nu g_\bot^{\mu\nu}, \ \  g_\bot^{\mu\nu}=g^{\mu\nu}-\frac12(n^\mu\nb^\nu+n^\nu\nb^\mu).
\eea
The  components of  $p_{hc}$ collinear to vector $\nb$ scale as
\bea
(p_{hc}\cdot n,p_{hc\perp}, p_{hc}\cdot \nb )\sim (1,\sqrt{\lambda},\lambda)m_c,\  p_{hc}^2\sim \lambda m_c^2,
\label{def:hcol2nb}
\eea 
and 
\bea
(p_{hc}\cdot n,p_{hc\perp}, p_{hc}\cdot \nb )\sim (\lambda,\sqrt{\lambda},1)m_c,\  p_{hc}^2\sim \lambda m_c^2,
\label{def:hcol2n}
\eea
for the momenta collinear to $n$. Remind, the we assume $\lambda\sim \Lambda/m_c$.  

The usoft momentum scales as $p^\mu_{us} \sim m_c v^2\sim \Lambda$ and describes low-energy nonpertubative modes.  Such correspondence  assumes that the small velocity $v$ scales as $v^2\sim \lambda$. This counting allows us to describe the soft overlap between NRQCD and collinear sectors consistently.  Notice,  that this correspondence is valid for the usoft modes only, and the counting for NRQCD matrix element like in Eq.(\ref{def:Rp21}), described by soft fields should not be converted into powers of $\lambda$.  

 A collinear momentum $p_c$ is  decomposed as 
\bea
p_c^\mu=(p_c\cdot n) \frac{\nb^\mu}{2}+ (p_c\cdot \nb) \frac{n^\mu}{2}+p_{c\bot} ^\mu,
\eea
The momentum components collinear to vector $\nb$ scale as
\bea
(p_c\cdot n, p_{c\perp}, p_c\cdot \nb )\sim (1,\lambda,\lambda^2)m_c,\  p_c^2\sim \lambda^2 m_c^2,
\label{def:col2nb}
\eea 
or   
\bea
(p_c\cdot n,p_{c\perp}, p_c\cdot \nb )\sim (\lambda^2,\lambda,1)m_c, \ p_c^2\sim \lambda^2 m_c^2,
\label{def:col2n}
\eea 
if they are associated with the sector $n$.

In order to simplify  formulas we also use short notation  $(p\cdot n)\equiv p_+$ and   $(p\cdot \nb)\equiv p_-$. 
It is also convenient to use the short notation for the transverse projection of Levi-Civita tensor
\bea
i\varepsilon_\bot[\mu,\nu]=\frac12 i\varepsilon_{\mu \nu \alpha \beta}n^\alpha\nb^\beta.
\eea

The collinear operators are defined in terms of the manifestly gauge-invariant  building blocs $\chi_{\nb}(x)$ and $\chi_n(x)$ describing particles with momenta scaling as  (\ref{def:col2nb}) and  (\ref{def:col2n}), respectively.  These blocs are defined as  
\bea
\chi_{\nb}(x)=W_{\nb}^\dagger (x)\nbn \psi_c(x), \   W_{\nb}^\dagger (x)=\text{P exp}\left( ig_s \int_{-\infty}^0 dt\ n\cdot A_c(x+t n) \right),
\eea
where $\psi_c$ and $A_c$ denote  collinear quark  and gluon fields, describing the particles with momenta (\ref{def:col2nb}).  The  field  $\chi_n(x)$ is defined analogously
\bea
\chi_{n}(x)=W_{n}^\dagger (x)\nnb \psi_c(x), \   W_{n}^\dagger (x)=\text{P exp}\left( ig_s \int_{-\infty}^0 dt\ \nb\cdot A_c(x+t \nb) \right),
\eea
with the collinear fields  corresponding to particles with momenta (\ref{def:col2n}). 

In order to simplify  expressions for the light-cone operators  the following notation is introduced
\bea
\chi_{\nb}(x_-)\equiv \chi_{\nb}(x_- n/2). 
\eea

Finally, let us add short remarks about the power counting of the decay amplitudes. The power counting for the collinear operators, which are discussed in the paper can be easily obtained  from the following formulas
\bea
 \chi_{n}\sim \chi_{\nb} \sim \lambda,\  \partial_\perp \chi_{\nb}\sim \partial_\perp \chi_{n}\sim \lambda^2, \  
 \partial_\perp \partial_\perp \chi_{\nb}\sim \partial_\perp \partial_\perp \chi_{n}\sim \lambda^3.
\eea

Using these rules one can easily find the counting for the collinear operators and their matrix elements. For the collinear matrix elements of twist-2 light-cone operator like in Eq.(\ref{Otw2}) this gives
\bea
\langle V(k', e_V)| \bar \chi_{n} (z_1)\Gamma \chi_{n}(z_2)|0\rangle \sim \lambda, 
\eea 
where it is taken into account that normalisation of the energetic vector particle gives the factor $\lambda^{-1}$.  Similarly, for the twist-3 and twist-4 valence  matrix elements in Eqs.(\ref{Otw3}) and (\ref{Otw4}) one finds
\bea
\langle V(k') \vert  \bar \chi_n (z_1) \Gamma \partial_\bot \chi_n(z_2)  \vert 0\rangle \sim \lambda^2,
\\
\langle V(k') \vert  \bar \chi_n (z_1) \Dsl{\bar{n}}\gamma_{\perp}^ \sigma \partial^\alpha_{\bot} \partial^{\beta}_\bot  \chi_n(z_2)  \vert 0\rangle \sim \lambda^3.
\eea 
These estimates allow one to get the power counting for the colour-singlet contributions in various decay amplitudes. Taking into account, that the NRQCD matrix element in Eq.(\ref{def:Rp21}) scales as $f_\chi\sim v^4$ one finds
\bea
A_{0,2}^{(LL)}\sim A_{2t}^{(TT)}&\sim &f_\chi\ \phi_{2V}*T* \phi_{2V} \sim v^4 \lambda^2, 
\\  
A_{1,2}^{(LT)}&\sim & f_\chi\ \phi_{2V}*T* \phi_{3V} \sim v^4 \lambda^3  ,  
\\
A_{0,2s}^{(TT)} &\sim&   f_\chi\ \phi_{3V}*T* \phi_{3V} + f_\chi\ \phi_{2V}*T* \phi_{4V}   \sim v^4 \lambda^4.
\eea
These results can also be obtained  from the explicit expressions in Eqs.(\ref{Ai**}), (\ref{ALTcoll}) and (\ref{ATTcols}) taking into accounts that $f_V\sim f^\bot_V\sim m_V\sim \lambda m_c$.  The colour-octet contributions scale as the appropriate colour-singlet ones because the soft overlap is logarithmic as it follows from the structure of the endpoint divergencies.

\section{Definitions and models of LCDAs }
\label{def:LCDA}
In this section we provide the definitions of the valence  LCDAs and their relationships with the  LCDAs defined in the literature, see {\it e.g.} refs.\cite{Ball:1998sk,Ball:1998ff,Ball:2003sc,Ball:2007rt}.

The twist-2  matrix elements in our notations read \footnote{For simplicity,  we skip in this section  the index $V$ for the polarisation vector: $(e^*_{V})_\mu\equiv e^*_\mu$ }
\begin{equation}
\left\langle V(k,e)\right\vert \bar{\chi}_{\nb}(x_{-})\Dsl n \chi_{\nb}(y_{-})\left\vert
0\right\rangle =-if_{V}m_{V}\ e_{+}^{\ast}\text{FT}\left[  \phi^{\Vert}_{2V} \right]  ,
\end{equation}%
\begin{equation}
\left\langle V(k,e)\right\vert \bar{\chi}_{\nb}(x_{-})\Dsl n \gamma_{\bot\nu}%
\chi_{\nb}(y_{-})\left\vert 0\right\rangle =-if_{V} k_{+} e_{\bot\nu}^{\ast} %
\text{FT}\left[  \phi^{\bot}_{2V}\right] ,
\end{equation}
where Fourier transformations FT in the {\it rhs} is defined as 
\begin{equation}
\text{FT}[\phi] =\int_0^1 du\  e^{i(k_{1}\cdot x)+i(k_{2}\cdot y)}\phi(u), \quad k_1=u k_+\bar n/2,\ k_2=\bar u k_+\bar n/2.
\label{def:FT}
\end{equation}%
  For numerical estimates we use the LCDA  models from ref.\cite{Ball:2007rt}  
  \begin{equation}
\phi_{2V}^{\Vert,\bot}(x)=6x\bar x\left(  1+a_{1V}^{\Vert,\bot}%
C_{1}^{3/2}(2x-1)+a_{2V}^{\Vert,\bot}C_{2}^{3/2}(2x-1)\right)  ,
\label{def:tw2mod}
\end{equation}
where the values of the moments $a_{nV}^{\Vert,\bot}$ are given in Table~\ref{app_LCDA}.
\begin{table}[h!]
\centering
\caption{ The values of the moments $a_{nV}^{\Vert,\bot}$  for different  LCDAs at $ \mu=2\ $GeV obtained in ref.~\cite{Ball:2007rt} . }
\label{app_LCDA}
\begin{tabular}
[c]{|c|c|c|c|}\hline
$V$  & $\omega$ & $\phi$ & $K^{\ast}$\\\hline
$f_{V}^{\Vert},$MeV  & $187(5)$ & $215(5)$ & $220(5)$\\\hline
$a_{1V}^{\Vert}$ &  $0$ & $0$ & $0.02(2)$\\\hline
$a_{2V}^{\Vert}$ &  $0.10(5)$ & $0.13(6)$ & $0.08(6)$\\\hline
$f_{V}^{\bot},$MeV & $151(9)$ & $186(9)$ & $185(10)$\\\hline
$a_{1V}^{\bot}$ & $0$ & $0$ & $0.03(3)$\\\hline
$a_{2V}^{\bot}$ &  $0.11(5)$ & $0.11(5)$ & $0.08(6)$\\\hline
\end{tabular}
\end{table}

The twist-3 valence matrix elements are defined as 
\bea
\left\langle V(k,e)\right\vert \bar{\chi}_{\nb}(x_{-})\Dsl n
{\partial}_{\bot\alpha}\chi_{\nb}(y_{-})\left\vert 0\right\rangle
= e_{\bot\alpha}^{\ast}\ k_{+}f_{V}m_{V}\ \text{FT}\left[
G^{\bot}_{3V}\right] ,
\\
\left\langle V(k,e)\right\vert \bar{\chi}_{\nb}(x_{-})\Dsl n \gamma_{5}{\partial}_{\bot\alpha}\chi_{\nb}(y_{-})\left\vert 0\right\rangle
=i\varepsilon_\perp [ e^{\ast}\alpha]\ k_{+}f_{V}m_{V}~\ \text{FT}\left[
\tilde G^{\bot}_{3V}\right]  ,
\\
\left\langle V(k,e)\right\vert \bar{\chi}_{\nb}(x_{-})\Dsl n
{\Dsl \partial }_{\bot}\chi_{\nb}(y_{-})\left\vert 0\right\rangle
=e_{+}^{\ast}\ f_{V}^{\bot}m_{V}^{2}\ \text{FT}\left[  G^{\Vert}_{3V}\right]  .
\eea
The relations of these LCDAs with ones in refs. \cite{Ball:1998sk,Ball:2007rt}  can be easily  obtained  rewriting the Lorentz covariant light-cone operators in the definitions  of the matrix elements in  \cite{Ball:1998sk,Ball:2007rt}  through the SCET fields $\chi_{\nb}$.  Neglecting the quark-gluon contributions we obtained the following expressions 
\bea
G_{3V}^{\bot}(u)&=&\frac{\bar{u}}{2}\int_{0}^{u}dv\ \frac{\phi^{\Vert}_{2V}(v)}{1-v}%
-\frac{u}{2}\int_{u}^{1}dv\frac{\phi^{\Vert}_{2V}(v)}{v}
\nonumber \\  
&&-\delta_{VK^*}\frac12 \frac{f_{K^{\ast}}^{\bot}}{f_{K^{\ast}}}\frac{m_{s}}{m_{K^*}}  % \tilde{G}_{\bot}^{(v)}(u)
\int_{0}^{u}dv %~\tilde{g}_{\bot}^{(v)}
\left( \frac{ \phi^{\bot}_{2K^*}(v) } {v}-\int_v^1dv'\frac{1}{v^{\prime 2}}\phi^{\bot}_{2K^*}(v') \right).
\label{G3Vbot}
\eea
\bea
\tilde G_{3V}^{\bot}(u)&=&\frac{\bar{u}}{2}\int_{0}^{u}dv\ \frac{\phi^\Vert_{2V}(v)}{1-v}
+\frac{u}{2}\int_{u}^{1}dv \frac{\phi^\Vert_{2V}(v)}{v}
%\nonumber \\ &&
 -\delta_{VK^*} \frac{f_{K^{\ast}}^{\bot}}{f_{K^{\ast}}}\frac{m_{s}}{m_{K^*}}\  \frac u2 \int_{u}^{1}dv\frac{\phi^{\bot}_{2K^*}(v)}{v^{2}}.
\eea
\bea
G^{\Vert}_{3V}(u)&=&4u\bar{u}\left[  \int_{0}^{u}dv\frac{\phi^{\bot}_{2V}(v)}%
{1-v}-\int_{u}^{1}dv\frac{\phi^{\bot}_{2V}(v)}{v}\right]
-\delta_{VK^*} \frac{f_{K^{\ast}}}{f_{K^{\ast}}^{\bot}}\frac{m_{s}}{m_{K^*}}
 \nonumber \\
&&\times 
\left ( 
4\int_{0}^{u}dv\ \left(  F(v)-\phi^\Vert_{2K^{\ast}} (v)\right) 
 +4u\bar{u}
\left[ 
 \int_{0}^u dv\frac{F(v)}{1-v}-\int_{u}^{1}dv\frac{F(v)}{v}
\right]
\right),
\label{G3VVert}
\eea
with
\begin{equation}
F(v)=\frac{1}{2}\left(  \phi^\Vert_{2K^{\ast}}(v)+\bar{v}\frac{d}{dv}\phi^\Vert_{2K^{\ast}}(v)\right)  .
\end{equation}
The symbol $\delta_{VK^*}$ is the Kronecker delta with respect to symbol  $V$.

The twist-4 valence matrix element is defined as 
\bea
\left\langle V(k,e)\right\vert
 \bar{\chi}_{\nb}(x_{-})\Dsl n \gamma_{\bot}^{\sigma}  \partial_{\bot}^\alpha \partial_{\bot}^\beta\bar{\chi}_{\nb}(y_{-})  
 \left\vert0\right\rangle 
 &=& -ik_{+}  f_{V}^{\bot}m_V^{2}\frac{1}{8}
 \left\{
(e_{\bot}^{\ast})^\sigma g_{\bot}^{\alpha\beta}\text{FT}[G^{(0)}_{4V}]\right.
\nonumber \\
&& \left. +(e_{\bot}^{\ast})_\rho G_{\bot}^{\alpha\beta\sigma\rho} \text{FT} [G^{(2)}_{4V}]
\right\}  ,
\label{def:G4}
\eea
where we use short notation
\begin{equation}
G_{\bot}^{\alpha\beta\sigma\rho}=g_{\bot}^{\alpha\sigma}g_{\bot}^{\beta\rho
}+g_{\bot}^{\beta\sigma}g_{\bot}^{\alpha\rho}-g_{\bot}^{\sigma\rho}g_{\bot
}^{\alpha\beta}.
\end{equation}

The first term in Eq.(\ref{def:G4}) is irrelevant for the calculations of  amplitudes $A_{0,2s}^{(TT)}$  because this LCDA describes the quark-antiquark pair with $L_z=0$ and therefore corresponding trace vanishes. Hence,  we only need to derive the analytical expression for LCDA $G^{(2)}_{4V}$.  In order to do this we proceed as follows. 

At  first step we use the  light-cone matrix element \cite{Ball:1998ff} 
\bea
\left\langle V(k,e)\right\vert \bar{\psi}(z_-)W_{\bar{n}}(z_-)\sigma^{\mu\nu}W^\dag_{\bar{n}}(-z_-)\psi(-z_-)\left\vert
0\right\rangle &= &-f_{V}^{\bot}\left(  e^*_{\bot\mu}k_{\nu}-e^*_{\bot\nu}k_{\mu}\right)  \text{FT}\left[  \phi^\bot_{2V}(u)\right]
\nonumber  %\\ &&
\eea
\\[-15mm]
\bea 
&&\phantom{123456789abcdef}
-f_{V}^{\bot}m^{2}_V \left(  k_{\mu}z_{\nu}-k_{\nu}z_{\mu}\right)  \frac
{(e^*z)}{(kz)^{2}}\text{FT}\left[  h_{\Vert}^{(t)}(u)\right]  
\nonumber \\ && \phantom{123456789abcdef}
-f_{V}^{\bot}m^{2}_V\frac{1}{2}\left(  e^*_{\bot\mu}z_{\nu}-e^*_{\bot\nu}z_{\mu}\right)
\frac{1}{(kz)}\text{FT}\left[  h_{3}(u)\right] . 
\label{def:lcm}
\eea

We rewrite  the operator in the {\it lhs} in terms of the valence operators using identity
\bea
W^\dag_{\bar{n}}(z_-)\psi(z_-)\simeq \chi_{\nb} (z_-)-\frac{\Dsl n}2(in\partial)^{-1} i\Dsl\partial_\bot \chi_{\nb} (z_-),
\label{psi-chi}
\eea
i.e. at this step  we neglect all quark-gluon operators. This gives the relation
\bea
G^{(2)}_{4V}(u)= 2u\bar u h^{(q)}_3(u).
\label{G4Vh3}
\eea  
The LCDA $h^{(q)}_3$ in (\ref{G4Vh3}) is a part of  the total LCDA
 \bea
 h_3(u)=h_3^{(q)}(u)+h_3^{(g)}(u),
 \eea 
 where  $h_3^{(g)}$ denotes the contribution of the quark-gluon LCDAs.  Notice that $h_3^{(q)}$ must satisfy the same normalisation as the $h_3$
 \bea
 \int_0^1 h_3(u)= \int_0^1 h^{(q)}_3(u)=1,
 \eea
 which ensures  the correct local normalisation of Eq.(\ref{def:lcm})  when  $h_3^{(g)}$ is neglected. 
 
 In order to derive  $h_3^{(q)}$  we consider the following correlation function  ($x^2\neq 0$)
 \bea
 \left\langle V(k,e) \right\vert \bar{\psi}(x)[x,-x]\sigma^{\mu\nu}\psi(-x)\left\vert 0 \right\rangle 
 =if_{V}^{\bot}\left( e^*_{\nu}k_{\mu} -e^*_{\mu}k_{\nu}\right)  \text{FT}\left[  \phi^\bot_{2V}
 +\frac{1}{4}m_V^{2} x^{2}A\right] 
 \nonumber
\\
+if_{V}^{\bot}m^{2}_V\left( k_{\nu}x_{\mu}- k_{\mu}x_{\nu}\right)  
\frac{(e^*x)}{(kx)^{2}}\text{FT}\left[ B\right]
 \nonumber
\\
 + if_{V}^{\bot}m^{2}_V\frac{1}{2}\left(e^*_{\nu}x_{\mu}-  e^*_{\mu}x_{\nu}\right)  \frac{1}{(kx)}\text{FT}\left[  C\right]  ,
 \label{def:cf}
 \eea  
 where   $[x,y]=\text P\exp\left\{ ig\int_0^1dt (x-y)_\mu A^\mu (tx+(1-t)y) \right\}$  and 
\begin{equation}
B=h_{\Vert}^{(t)}-\frac{1}{2}\phi^\bot_{2V}-\frac{1}{2}h_{3},\ \ \ C=h_{3}-\phi^\bot_{2V}.
\end{equation}
 The twist-3 LCDA $h_{\Vert}^{(t)}$  is given by \cite{Ball:1998sk}
 \bea
h_{\Vert}^{(t)}(u)=(2u-1)\left(  \int_{0}^{u}dv\frac{\phi^\bot_{2V}\left(  v\right)
}{1-v}-\int_{u}^{1}dv\frac{\phi^\bot_{2V}\left(  v\right)  }{v}\right)+\dots,
\eea
 where dots denote contributions of the quark-gluon operators. 
 Expanding the {\it  lhs} of (\ref{def:cf}) with the help of Eq.(\ref{psi-chi})  and the {\it rhs}  up to twist-4 accuracy  one finds the linear  relation, which involves $h^{(q)}_3$ and  $G^{(2)}_{4V}$. Excluding  $G^{(2)}_{4V}$ with the help of Eq.(\ref{G4Vh3}) yields the following  differential equation for the $ h_{3}^{(q)}$
\begin{equation}
2\left(  2h_{\Vert}^{(t)}-\phi^\bot_{2V}-h^{(q)}_{3}\right)(u)  =\left[  u\bar{u}%
h^{(q)}_{3}(u)\right]  ^{\prime\prime},~\label{h3deq}%
\end{equation}
Using the twist-2 LCDA from Eq.(\ref{def:tw2mod}) one finds the solution of this equation for $h_{3}^{(q)}$    
 \begin{equation}
 h_{3}^{(q)}(u)=6  u\bar{u} \left(
1+\frac{a_{1}^{\bot}}3C_{1}^{3/2}(2u-1)+\frac{a_{2}^{\bot}}{6}C_{2}^{3/2}(2u-1)\right)  .
\label{res:h3q}
\end{equation}
 This solution  agrees with the pure quark contribution for $h_3$ in case of $\rho$-meson, which was derived in ref.~\cite{Ball:1998ff}  using a different technical method.

\section{ Projection operators for  collinear matrix elements}
\label{def:Projections}
The projections of the higher-twist light-cone matrix elements are often derived  using the covariant parametrisation of the  correlation function, see {\it e.g.} discussion ref.~\cite{Beneke:2000wa}.  In this work we use a different technique, which is based on the light-cone cone expansion of the external  collinear fields  to a given accuracy  and reduction of the light-cone operators to the valence ones.  Such approach  involves the QCD equation of motions in order to integrate the small components of the collinear fields and  allows one to deal  with the light-cone operators only.  This approach has many advantages, in particular, one can apply  the field redefinition in order to get  the collinear Wilson lines in the operators and therefore better maintain the colour gauge invariance.  The valence LCDAs have simple analytical  properties, which  simplify the analysis  of  infrared singularities.  The technical details was described in ref.~\cite{Kivel:2019wjh} and we will not repeat them  here.  
 
  In the following  we just provide the analytical expressions for the different projection operators $M_i$, which are used for the calculation of the colour-singlet contributions in the collinear factorisation framework, see Eq.(\ref{calc:AJLT}).   Below we assume that these   projection operators are associated with the off light-cone matrix element
 \bea
  \left\langle V(k,e)\right\vert \bar{\chi}_{\alpha_1}(x)\ \chi_{\alpha_2}(y)\left\vert 0\right\rangle \to  \int_0^1 du
  \sum_{i=2,3,4}[\hat M_2(u,k) ]_{\alpha_2 \alpha_1}\  e^{i(k_1x)+i(k_2y)}
  \label{def:Mi}
  \eea
  where we  assume that $x^2\neq 0$ and $y^2\neq 0$ and we also skip the colour structure for simplicity.  
  In the following expressions we assume   
  \bea
  k\equiv  2m_c \frac{\bar n}{2}, \quad  k'\equiv  2m_c \frac{n}{2}.
  \eea
  We also write all Dirac matrices in the square brackets $[\dots]$ and do not show the indexes  $\alpha_i $ explicitly. Some  terms  in $M_i$  do not depend on  transverse derivatives $\partial / \partial k_\bot$ then  the momenta $k_i$  in 
  $e^{i(k_1x)+i(k_2y)}$ in Eq.(\ref{def:Mi}) are  defined as in Eq.(\ref{def:FT}). For the terms with  transverse derivatives the momenta $k_i$  must be modified  as 
 \bea
 \quad k_1=u k_+\bar n/2+k_\bot,\ k_2=\bar u k_+\bar n/2-k_\bot.
 \eea 
 
 The simplest twist-2 projection reads
 \bea
 \hat M_2(u,k)=-if_{V}m_{V}\frac{\left(  k^{\prime}\cdot e^{\ast}\right)
}{(kk^{\prime})}\frac{1}{4}\left[  \Dsl k\right] \ \phi^\Vert_{2V}(u)+if_{V}^{\bot}\frac{1}{4}\left[  \Dsl k\Dsl e_{\bot}^{\ast}\right] \  \phi^{\bot}_{2V}(u).
 \eea 
 
 The twist-3 projection can be written as 
  \bea
 \hat M_3(u,k)= if_{V}m_{V} \left(  e_{\bot}^{\ast}\right) ^{\alpha} \hat M^{\bot }_{3\alpha}(u,k)
 + if_{V}^{\bot}m_{V}^{2}\frac{\left(  e^{\ast}\cdot k^{\prime}\right)  }{(k\cdot k^{\prime})} \hat M^\Vert_3(u, k),
 \eea
 where 
 \bea
 \hat M^{\bot}_{3\alpha }(u,k)&=&\left( G^{\bot}_{3V}(u)g^\bot_{\alpha\beta}+  \tilde G^{\bot}_{3V}(u)
i\varepsilon^\perp_ {\alpha\beta } [\gamma_5]  \right)
%\nonumber \\ &&\times 
\frac{1}{4} \left( \left[  \Dsl k\right]  \frac{\partial}{\partial k_{\bot\beta}}  
+ \frac{\left[  \Dsl k^{\prime}\gamma_{\bot}^\beta\Dsl k\right]}{2\bar u(kk^{\prime})}
 - \frac{\left[  \Dsl k \gamma_{\bot}^\beta k^{\prime}\right] }{2u(kk^{\prime})}
\right)
\nonumber \\
 &+& \frac{f_{V}^{\bot}}{f_{V}}  \frac{m_{q}}{m_V} \frac{1}{2(kk^{\prime})} \frac{1}{4}  
 \left[ \Dsl k\Dsl k^{\prime} \gamma_\bot^\alpha \right]  
  \frac{\phi^{\bot}_{2V}(u)}{u}
  - \frac{f_{V}^{\bot}}{f_{V}} \frac{m_{\bar q}}{m_V}  \frac{1}{2(kk^{\prime})} \frac{1}{4}  
 \left[ \Dsl k^{\prime}\Dsl k \gamma_\bot^\alpha \right]  
  \frac{\phi^{\bot}_{2V}(u)}{1-u}
  \eea
 \bea
 \hat M^{\Vert}_3(u,k)&=& G^{\Vert}_{3V}(u)\frac{1}{4}
 \left( 
  -\frac{1}{2}\left[  \Dsl k \gamma_{\bot}^{\alpha}\right] \frac{\partial}{\partial k_{\bot}^{\alpha}} 
+ 2\frac{\left[ \Dsl k^{\prime}\Dsl k\right]  }{\bar u (kk^{\prime})}
+2\frac{\left[ \Dsl k \Dsl k^{\prime} \right]  }{u(kk^{\prime})}   
 \right)
 \nonumber \\  
 &+& \frac{f_{V}}{f_{V}^{\bot}}   \frac{m_{q}}{m_V} \frac{1}{2(kk^{\prime})}
  \frac{1}{4}\left[\Dsl k\Dsl k' \right]  \frac{\phi_{2V}^\Vert(u)}{u}
  +\frac{f_{V}}{f_{V}^{\bot}}   \frac{m_{\bar q}}{m_V} \frac{1}{2(kk^{\prime})}
  \frac{1}{4}\left[\Dsl k'\Dsl k \right]  \frac{\phi_{2V}^\Vert(u)}{1-u}.
  \eea 
 The terms with  quark(antiquark) masses $m_{q(\bar q)}$  are only considered for  $K^*$-mesons in order to get  strange mass corrections when $m_q=m_s$ or $m_{\bar q}=m_s$.

The twist-4 projection is given by 
\begin{equation}
\hat M_4(u,k)=if_{V}^{\bot} m_{V}^{2}\frac{1}{8}
\left( G^{(2)}_{4V}(u) M^{(2)}_{4}(u,k)  +G^{(0)}_{4V}(u) M^{(0)}_{4}(u,k) \right),
\end{equation}
where  
\bea
M^{(0)}_{4}(u,k)&=&-\frac{1}{4}\left[  \Dsl k \Dsl e _{\bot}^{\ast}\right]
  \frac{1}{2} \frac{\partial}{\partial k_{\bot\alpha}}
  \frac{\partial}{\partial k_{\bot}^{\alpha}} 
   -\frac{1}{4}  \left[\Dsl k \Dsl e _{\bot}^{\ast}\right] \frac{1}{2(kk^{\prime})}
  \left(\frac{1}{u}-\frac{1}{\bar u}\right)  k^{\prime\alpha}\frac{\partial}{\partial l^{\alpha}}
 \nonumber  \\
  &&
  +\frac{1}{4} 
  \left( 
   \frac{1}{\bar u} \left[  \Dsl k ^{\prime}\gamma_{\bot}^{\alpha}\Dsl k \Dsl e _{\bot}^{\ast}\right]
 -\frac{1}{u} \left[  \Dsl k \Dsl e _{\bot}^{\ast}\gamma_{\bot}^{\alpha}\Dsl k^{\prime}\right]
 \right)
   \frac{1}{2(kk^{\prime})}\frac{\partial}{\partial k_{\bot}^{\alpha}}.
 \eea
 The auxiliary momentum $l$  for the term with derivative $\partial/\partial l^{\alpha}$ is defined through the momenta $k_i$ in  Eq.(\ref{def:Mi})  as     
 \bea
k_1=u k_+\bar n/2+l,\ k_2=\bar u k_+\bar n/2-l.
 \eea 
 \bea
  \hat M^{(2)}_{4}(u,k)&=& \frac{1}{4}\left[  \Dsl k \gamma_{\bot}^{\sigma}\right]
(e_{\bot}^{\ast})^\rho\left\{  \frac{1}{2}g^{\bot}_{\sigma\rho}
\frac{\partial}{\partial k_{\alpha\bot}}\frac{\partial}{\partial k_{\bot}^{\alpha}}
 -\frac{\partial}{\partial k_{\bot}^\sigma}\frac{\partial}{\partial k_{\bot}^\rho}\right\} 
\nonumber \\
&&  +
\frac{1}{2(kk^{\prime})}
 \frac{1}{4} \left(
 \frac{1}{\bar u} \left[ \Dsl k^{\prime}\gamma_{\bot\beta} \Dsl k\gamma_{\bot\sigma}\right] 
+\frac{1}{u} \left[  \Dsl k\gamma_{\bot\sigma}\Dsl k^{\prime} \gamma_{\bot\beta}\right] 
\right)
 (e_{\bot}^{\ast})_\rho G_{\bot}^{\alpha\beta\sigma\rho}
 \frac{\partial}{\partial k_{\bot}^{\alpha}}
\nonumber\\
&&  +\frac{2}{(kk^{\prime})} \frac{1}{u \bar u} \frac{1}{4}\left[  \Dsl e _{\bot}^{\ast} \Dsl  k^{\prime}\right] .
 \eea

\acknowledgments
I am grateful to Qi Huang and the other authors of ref.~\cite{Huang:2021kfm} for kindly clarifying some aspects of their work.

%\paragraph{Note added.} This is also a good position for notes added after the paper has been written.

\end{document}